\documentclass[aps,prc,showkeys,nofootinbib]{revtex4}

\usepackage{graphicx}
\usepackage{dcolumn}
\usepackage{amsthm}
\usepackage{bm}

\usepackage{isotope}
\usepackage{physics}

\newcommand{\xibf}{\mbox{\boldmath $\xi$}}
\newcommand{\rhobf}{\mbox{\boldmath $\rho$}}

\newcommand{\varepsilonbf}{\mbox{\boldmath $\varepsilon$}}
\newcommand{\pibf}{\mbox{\boldmath $\pi$}}

\begin{document}

\title{Systematic study of bremsstrahlung emission in reactions with light nuclei in cluster models
}

\author{S.~P.~Maydanyuk$^{1,2}$}\email{sergei.maydanyuk@wigner.hu}%
\author{V.~S.~Vasilevsky$^{3}$}\email{vsvasilevsky@gmail.com}%

\affiliation{$^{(1)}$Wigner Research Centre for Physics, Budapest, 1121, Hungary}
\affiliation{$^{(2)}$Institute for Nuclear Research, National Academy of Sciences of Ukraine, Kyiv, 03680, Ukraine}
\affiliation{$^{(3)}$Bogolyubov Institute for Theoretical Physics, Metrolohichna str., 14b, Kyiv, 03143, Ukraine}

\date{\small\today}

\begin{abstract}
A new model of bremsstrahlung emission in the scattering of light nuclei is constructed with main focus on strict cluster formulation of nuclear processes.
Analysis is performed in frameworks of the folding approximation of the formalism with participation of $s$-nuclei.
Reactions
$p + \isotope[4]{He}$,
$\isotope[2]{D} + \isotope[4]{He}$,
$\isotope[3]{H} + \isotope[4]{He}$,
$\isotope[3]{He} + \isotope[4]{He}$
are included to analysis.
Systematic analysis of properties of emission of bremsstrahlung photons in the wide region of kinetic energy of relative motion of two nuclei from 7 to 1000~MeV is performed.
Influence of the oscillator length on the calculated spectra of bremsstrahlung emission is analyzed.
On the example of $\isotope[3]{H} + \isotope[4]{He}$,
dependence of the bremsstrahlung spectra on parameters of nuclear component of interacting potential is established (at first time for the light nuclei).
Experimental bremsstrahlung data for the proton-deuteron scattering and proton-$\alpha$-particle scattering are analyzed on the basis of this model.
\end{abstract}



\maketitle

\section{Introduction
\label{sec.introduction}}

The bremsstrahlung emission of photons accompanying nuclear reactions is an important topic of nuclear physics,
it has been causing essential interest for from many researchers long time
(see reviews~\cite{Amusia.1988.PhysRep,Pluiko.1987.PEPAN,Kamanin.1989.PEPAN,Bertulani.1988.PhysRep}).
This is explained by that the spectra of bremsstrahlung photons are calculated on the basis of nuclear models with included mechanisms of reactions, interactions between nuclei, dynamics, many other physical issues.
Measurements of these photons and their analysis provide information about all these aspects, verifying suitability of the developed models.

Investigations of bremsstrahlung emission in the proton-nucleus scattering have shown important role of incoherent bremsstrahlung processes.
In particular, in Ref.~\cite{Maydanyuk.2012.PRC,Maydanyuk_Zhang.2015.PRC} (see also Ref.~\cite{Liu_Maydanyuk_Zhang_Liu.2019.PRC.hypernuclei})
a formalism which takes into account both coherent and incoherent processes has been formulated and
it was found that the incoherent contribution is essentially larger than the coherent one in the full bremsstrahlung emission.
Moreover, inclusion of the incoherent processes to formalism improves agreement between calculated cross section and experimental data~\cite{Goethem.2002.PRL} for
$p + \isotope[197]{Au}$ at energy of proton beam $E_{\rm p}=190$~MeV.
Another useful advance of incoherent bremsstrahlung is explanation of plateau in the middle part of experimental cross section\cite{Goethem.2002.PRL},
while coherent contribution gives only logarithmic behavior of the calculated cross section that is not enough for good description of the data~\cite{Goethem.2002.PRL}.
In such a formalism, full operator of emission of bremsstrahlung photons can be explicitly separated on two groups of terms.
One group (coherent bremsstrahlung) includes terms with momentum defined on relative distance between center-of-mass of the nucleus-target and the scattered proton
(for example, for the $\alpha$-nucleus scattering this is Eq.~(B10) in Ref.~\cite{Liu_Maydanyuk_Zhang_Liu.2019.PRC.hypernuclei}).
The second group (incoherent bremsstrahlung) includes rest of terms without momentum from relative distance between the nucleus-target and the scattered proton.
But it includes momenta of relative distances between individual nucleons of the nucleus-target and the scattering proton
(for the $\alpha$-nucleus scattering this is Eq.~(B11) with addition (B12) in Ref.~\cite{Liu_Maydanyuk_Zhang_Liu.2019.PRC.hypernuclei}).
By simple words, incoherent bremsstrahlung has origin from many nucleon dynamics, while coherent bremsstrahlung is related with two body (proton-nucleus) dynamics.

Thus, consideration of nuclear scattering as many-nucleon quantum mechanical problem allows one to increase essentially accuracy of description of bremsstrahlung spectra.
Another important message from those investigations is understanding about important role of magnetic moments of nucleons in nuclei in reactions (i.e. magnetic emission based on magnetic moments of nucleons in nuclei is more intensive essentially than electric emission based on electric charges of nucleons in nuclei).
Attempts to determine accurately relations between parameters of individual nucleons in the studied nuclear process and emission of bremsstrahlung photons give more deep understanding about nuclear (nucleon-nucleon) interactions.
This increases motivation to construct the fully cluster formalism for the nucleon-nucleus and nucleus-nucleus scattering in the bremsstrahlung problem.

In these regards, one can remind important investigations
of bremsstrahlung emission in reactions with light nuclei on the basis of cluster models.
Such investigations can reveal new information about properties of wave functions of continuous cross section
states of two colliding nuclei. This process is complementary to the capture
reaction or photodisintegration. However, contrary to these two processes,
the bremsstrahlung is more complicated from numerical point of view as a
calculation of the cross section involves two wave function of states in continuous
energy region.
This process has been investigated within microscopic two-cluster models~%
\cite{
1985NuPhA.443..302B,1990PhRvC..41.1401L,1990PhRvC..42.1895L,%
1991NuPhA.529..467B,1991PhRvC..44.1695L,1992NuPhA.550..250B,Liu.1992.FBS,%
1993nuco.conf..423K,%
Dohet-Eraly.2013.PRC,2013JPhCS.436a2030D,%
Dohet-Eraly.2013.PhD,2014PhRvC..89b4617D,2014PhRvC..90c4611D}
[see also Ref.~\cite{Baye.2012.PRC86}].
In Tabl.~\ref{table.1} we collect all available information about such investigations.
%
\begin{table}
\begin{center}
\begin{tabular}{|c|c|c|c|} \hline
  Nucleus & Clusterization & Papers \\ \hline
  \isotope[8]{Be} & $\alpha + \alpha$ &
    \cite{1985NuPhA.443..302B}, \cite{1991NuPhA.529..467B},
    \cite{1992NuPhA.550..250B},
    \cite{Liu.1992.FBS},
    \cite{1993nuco.conf..423K},
    \cite{2013JPhCS.436a2030D},
    \cite{Dohet-Eraly.2013.PhD},
    \cite{2014PhRvC..90c4611D}
    \\
  \isotope[7]{Be} & $\alpha + \isotope[3]{He}$ &
    \cite{1990PhRvC..41.1401L}, \cite{1991PhRvC..44.1695L} \\
  \isotope[5]{Li} & $\alpha + p$ &
    \cite{1990PhRvC..42.1895L}, \cite{1991PhRvC..44.1695L},
    \cite{Dohet-Eraly.2013.PhD}, \cite{2014PhRvC..89b4617D} \\ \hline
\end{tabular}
\end{center}
\caption{Nuclei and clusterizations
used in investigations of bremsstrahlung in the nuclear scattering on the basis of microscopic two-cluster models.
}
\label{table.1}
\end{table}

The first investigation has been performed in Ref.~\cite{1985NuPhA.443..302B}
where interaction of two $\alpha$ particles were considered. Theoretical data
was obtained for the initial energy of interacting $\alpha$ particles $E_{i}\leq 10$~MeV.
Different geometry of the reaction has been discussed in
detail and optimal geometry of experiments were recommended.
In Ref.~\cite{2013JPhCS.436a2030D} the same model was used to study the bremsstrahlung in
$\alpha$-$\alpha$ collisions with a realistic nucleon-nucleon potential.
And a wider energy range of two $\alpha$ particles ($E_{i}\leq 50$~MeV) was analyzed.
The bremsstrahlung emission in the $\alpha +^{3}$He collision has been
studied in Ref.~\cite{1990PhRvC..41.1401L} within the resonating group
method. Contribution of narrow resonance states $7/2^{-}$ and 5/2$^{-}$ to
the bremsstrahlung \ cross sections are thoroughly studied.

The resonating group method has been emplied in Ref.~\cite{1990PhRvC..42.1895L}
to investigate the bremsstrahlung process in the
interaction of protons with an $\alpha$ particle. The model correctly
reproduced phase shift of the elastic $p + \alpha$ scattering and the
parameters of the 3/2$^{-}$ and $1/2^{-}$ resonance states in $^{5}$Li. A
good agreement between available experimental data and results of
calculations was achieved. The bremsstrahlung cross section has been
calculated for the energy of incident proton $0\leq E_{p}\leq 25$~MeV.

Note that there is essentially larger volume of experimental bremsstrahlung data obtained with higher precision,
which has not been studied by the cluster approaches above (ideology of cluster formalism).
These are data~\cite{Goethem.2002.PRL} for the proton-nucleus scattering
(these are data for $p + \isotope[197]{Au}$ at energy of proton beam $E_{\rm p} = 190$~MeV and
inside photon energy region $E_{\gamma} = 40$ -- 170~MeV,
see also Refs.~\cite{Edington.1966.NP,Koehler.1967.PRL,Kwato_Njock.1988.PLB,Pinston.1989.PLB,Pinston.1990.PLB}),
data~\cite{Boie.2007.PRL,Boie.2009.PhD} for $\alpha$ decay of \isotope[210]{Po}
(see also Refs.~\cite{D'Arrigo.1994.PHLTA,Kasagi.1997.JPHGB,Kasagi.1997.PRLTA,Maydanyuk.2008.EPJA,Giardina.2008.MPLA}
for nuclei \isotope[210]{Po}, \isotope[214]{Po}, \isotope[226]{Ra}, \isotope[244]{Cm}),
data~\cite{Eremin.2010.IJMPE,Maydanyuk.2010.PRC} for the spontaneous fission of heavy nuclei~\isotope[252]{Cf}
(see also Refs.~\cite{Ploeg.1995.PRC,Kasagi.1989.JPSJ,Luke.1991.PRC,Varlachev.2007.BRASP,Hofman.1993.PRC,Pandit.2010.PLB}).
There are experimental investigations of dipole $\gamma$-ray emission with incident energy
in the $\isotope[32]{S} + \isotope[100]{Mo}$ and $\isotope[36]{S} + \isotope[96]{Mo}$ fusion reactions at $E_{\rm lab} =196$~MeV and 214.2~MeV, respectively,
with aim to probe evolution of fusion with incident energy
\cite{Pierroutsakou.2005.PRC}.
$\gamma$-decays of excited states in the energy region of the Pygmy dipole states in heavy nuclei have been observed~\cite{Kmiecik.Maj.2020.ActaPP}
(see also Ref.~\cite{Lanza.2023.ProgPartNuclPhys}).

One can note that the most of forces were put by researchers on study of bremsstrahlung in the proton-nucleus scattering and fission.
However, strict cluster model has not been constructed to describe those processes up to the level of good description of experimental data.
Some authors wrote in published papers about attractive perspective to realize this approach~\cite{Kopitin.1997.YF}.
Note on the microscopic models of $\alpha$ decay
\cite{Thomas.1954.PTP,Delion.1992.PRC,Xu.2006.PRC,Delion.2013.PRC,
Silisteanu.2012.ADNDT,Ivascu.1990.PEPAN,Lovas.1998.PRep,Hodgson.2003.PRep},
however emission of bremsstrahlung photons has not been studied on the basis of those approaches.
Also there are theoretical investigations on emission of bremsstrahlung during ternary fission of heavy nuclei \isotope[252]{Cf} \cite{Maydanyuk.2011.JPCS}.
As it was found in that paper, the bremsstrahlung spectra are essentially dependent on the different scenarios of dynamics of fission, that can be studied by means of bremsstrahlung analysis.
Properties of hypernuclei in the scattering have been studied in connection with analysis of bremsstrahlung emission~\cite{Liu_Maydanyuk_Zhang_Liu.2019.PRC.hypernuclei}.
However, those investigations above were performed without cluster basis.

Summarizing all issues described above, we see attractive perspective to construct such a formalism and to describe existed experimental information of bremsstrahlung in nuclear reactions in the unified way.
So, the first step in realization of this program is aim of this paper, where we put focus on scattering of nuclei with small number of nucleons
in frameworks of the folding approximation.

The paper is organized in the following way.
In Sec.~\ref{sec.cluster} a new cluster model of emission of bremsstrahlung photons in scattering of light nuclei is formulated.
In Sec.~\ref{sec.folding.1} the folding approximation of cluster model is described with main formulas for practical calculations for $s$-nuclei.
In Sec.~\ref{sec.analysis} properties of emission of bremsstrahlung photons for scattering are studied on the basis of the model presented above.
Here, we analyze
parameters which have essential influence on the accuracy of calculations of the spectra,
we calculate spectra in dependence on kinetic energy of relative motion between two nuclei in the wide energy region,
we estimate spectra for different nuclei at the same energy of relative motions between nuclei,
we analyze role of oscillator length of nuclei in calculations of the spectra,
we look for sensitivity of the shape of the spectra on the nuclear component of the interaction potential,
we describe experimental bremsstrahlung data for proton-deuteron scattering on the basis of the model.
Conclusions and perspectives are summarized in Sec.~\ref{sec.conclusions}.
Some useful details of the model are presented in Appendixes.
Here, we give
formalism and calculations of the matrix elements in the folding approximation (see Appendix~\ref{sec.app.folding}),
multiple expansion of the matrix elements of bremsstrahlung (see Appendix~\ref{sec.app.multipole}),
formalism of polarizations of the emitted photon (see Appendix~\ref{sec.app.polarization}),
calculation of angular integrals 
(see Appendix~\ref{sec.app.2}).

\section{Cluster formalism
\label{sec.cluster}}

To study the bremsstrahlung emission in interacting of light nuclei, we are
going to employ cluster models. First of all we need to specify the term
\emph{``cluster model''}. By saying cluster model we mean one of numerous realizations
(versions) of the resonating group method (RGM). The resonating group
method suggests that the properties of atomic nuclei and different types of
nuclear reactions can be described by assuming there are stable formations
of nucleons comprising clusters. A nucleon-nucleon interaction of nucleons
belonging to different clusters
creates a cluster-cluster interactions. Main difference of  various types
of the RGM consists (i)  in different shapes of wave functions describing
internal structure of clusters and (ii) in  different algorithms of solving
the many-particle Schr\"{o}dinger equation or equivalent effective two-body
equations of the RGM derived by J. Wheeler \cite{1937PhRv...52.1083W}, \cite%
{1937PhRv...52.1107W}.

Note that models which  consider clusters as structureless particles we
will call them as \emph{potential models}.

Now, we shall consider interaction between clusters, when antisymmetrization
is taken into account or neglected. More precisely, we employ totally
antisymmetric functions for a two--cluster configuration $A=A_{1}+A_{2}$
\begin{equation}
  \Psi ^{(A) }=
  \widehat{\mathcal{A}}\left\{ \Phi _{1}\left( A_{1}\right) \
  \Phi_{2}\left( A_{2}\right) \ \phi\,
  (\mathbf{r})
  \right\}
\label{eq:F00}
\end{equation}
and also the functions of so-called \emph{folding model}
\begin{equation}
  \Psi ^{(F)}=\Phi _{1}\left( A_{1}\right) \ \Phi _{2}\left( A_{2}\right) \
  \phi\,
   (\mathbf{r}),
\label{eq:F01}
\end{equation}
when antisymmetrization of nucleons of different cluster are neglected.
To describe relative motion of clusters, one can use distance between clusters $\mathbf{r}$
\begin{equation}
  \mathbf{r} =
  \left[ \displaystyle\frac{1}{A_{1}} \displaystyle\sum_{i\in A_{1}} \mathbf{r}_{i} -
  \displaystyle\frac{1}{A_{2}} \displaystyle\sum_{j\in A_{2}} \mathbf{r}_{j} \right].
\label{eq:F003}
\end{equation}
It is assumed, that the wave functions $\Phi _{i}\left( A_{i}\right) $,
describing internal structure of clusters, are translatoinally invariant and
antisymmetric ones. In Eq. (\ref{eq:F00}), the antisymmetrization operator $%
\widehat{\mathcal{A}}$ permutes nucleons between clusters and thus realize
the Pauli principle correctly. It is well-known, that the Pauli principle
play an important role especially for low-energy region of interacting
clusters. This approximation we will call the standard version of the
resonating group method (RGM). \ The second approximation which is presented
by wave function (\ref{eq:F01}) is called the Folding model (FM) or folding
approximations.

In what follows, wave functions  $\phi
\left( \mathbf{r}\right) $ are to represented in the spherical coordinates
\begin{equation}
  \phi \left( \mathbf{r}\right) =
  R_{E l}\left( r\right) Y_{lm} \left( \widehat{\mathbf{r}}\right),
\label{eq:F004}
\end{equation}
where $\widehat{\mathbf{r}} = \mathbf{r} / r$, $|\mathbf{r}| = r$.
Wave functions (\ref{eq:F00}) and (\ref{eq:F01}) suggest approximate
solutions for the Schr\"{o}dinger equation%
\begin{eqnarray}
\left( \widehat{H}-E\right) \Psi ^{(A)} &=&0,  \label{eq:F00A} \\
\left( \widehat{H}-E\right) \Psi ^{(F)} &=&0  \label{eq:F00B}
\end{eqnarray}
with a microscopic Hamiltonian $\widehat{H}$ which consists of the kinetic
energy operator in the center of mass motion and a sum of pairwise
nucleon-nucleon potentials.

By multiplying these equations from the left on the product $\Phi _{1}\left(
A_{1}\right) \ \Phi _{2}\left( A_{2}\right) $ and integrating over internal
spacial, spin and isospin coordinates of \ nucleons, we obtain
integro-differential equation \ for $\ \phi \left( \mathbf{q}\right) $ when
the Pauli principle is treated correctly, or differential equation when the
folding approximation is used. \ The later can be written as
\begin{equation}
  \left\{ -\frac{\hbar ^{2}}{2m_{N} \mu}\Delta _{\mathbf{r}}+\widehat{V}^{(F)}\left(
  \mathbf{r}\right) -E\right\} \phi \left( \mathbf{r}\right) =0,
\label{eq:F001}
\end{equation}
where $\mu$ is the reduced mass
\begin{equation}
\mu = \frac{A_{1}A_{2}}{A_{1}+A_{2}}
\end{equation}
and $m_{N}$ is mass of nucleon.
It is important to underline that the folding potential $\widehat{V}^{(F)}\left( \mathbf{r}\right)$
is a key component of a nonlocal
inter-cluster potential appeared in the standard version of the RGM. The
folding potential $\widehat{V}^{(F)}\left( \mathbf{r}\right) $ is totally
determined by the shape of nucleon-nucleon potential and density
distributions of nucleons in each cluster.

\subsection{Potential in the folding approximation
\label{sec.folding.potential}}

In the folding approximation, as it was pointed out above, the inter-cluster
potential is local and may be easily calculated, especially when simple
shell-model functions $\Phi _{i}(A_{i})$ are used to describe internal state
of clusters.

The folding potential is the integral:
\begin{equation}
  \widehat{V}^{(F)}(\mathbf{r}) =
  \displaystyle\sum_{i\in A_{1}} \displaystyle\sum_{j\in A_{2}}
  \displaystyle\int dV_{1}\, dV_{2}\; |\Phi (A_{1})|^{2}\,
  \widehat{V}(\mathbf{r}_{i} - \mathbf{r}_{j})\: |\Phi (A_{2})|^{2},
\label{eq:F02}
\end{equation}
where integration is performed over all coordinates
\begin{equation}
\begin{array}{llllll}
  dV_{1} = \displaystyle\prod\limits_{i\in A_{1}} d\mathbf{r}_{i}, &
  dV_{2} = \displaystyle\prod\limits_{i\in A_{2}} d\mathbf{r}_{i}.
\end{array}
\end{equation}
As wave functions $\Phi _{1}\left( A_{1}\right) $ and$\ \Phi _{2}\left(
A_{2}\right) $ are translationally invariant, they actually depends on
coordinates
\begin{eqnarray}
\mathbf{r}_{i}^{\prime } &=& \mathbf{r}_{i}-\mathbf{R}_{1},\quad \mathbf{R}%
_{1}= \displaystyle\frac{1}{A_{1}} \displaystyle\sum_{i\in A_{1}}\mathbf{r}_{i},\quad i\in A_{1},
\label{eq:F03} \\
\mathbf{r}_{j}^{\prime } &=&\mathbf{r}_{j}-\mathbf{R}_{2},\quad \mathbf{R}%
_{2} = \displaystyle\frac{1}{A_{2}} \displaystyle\sum_{j\in A_{2}}\mathbf{r}_{j},\quad j\in A_{2},
\nonumber
\end{eqnarray}
respectively. Thus we have to switch to these coordinates
%
\begin{equation}
\begin{array}{llllll}
  \widehat{V}^{(F)}(\mathbf{r}) = 
  \displaystyle\sum_{i\in A_{1}} \displaystyle\sum_{j\in A_{2}}
  \displaystyle\int\, dV_{1}^{^{\prime }}\; dV_{2}^{\prime }\;
  |\Phi (A_{1})|^{2}\, \widehat{V}(\mathbf{r}_{i}^{\prime } - \mathbf{r}_{j}^{\prime } + \mathbf{r})\, |\Phi (A_{2})|^{2},
\end{array}
\label{eq:F05}
\end{equation}
%
where
\begin{eqnarray}
  \mathbf{r} & = &
  \mathbf{R}_{1} - \mathbf{R}_{2} =
  \displaystyle\frac{1}{A_{1}} \displaystyle\sum_{i\in A_{1}} \mathbf{r}_{i} -
  \displaystyle\frac{1}{A_{2}} \displaystyle\sum_{j\in A_{2}} \mathbf{r}_{j}  \label{eq:F06} \\
  dV_{1}^{\prime } & = & \displaystyle\prod\limits_{i\in A_{1}} d\mathbf{r}_{i}^{\prime},\quad
  dV_{2}^{\prime } = \displaystyle\prod\limits_{j\in A_{2}}\, d\mathbf{r}_{j}^{\prime }
\nonumber
\end{eqnarray}
and $\Phi (A_{\alpha })$ is a many-particle shell model function, describing
internal motion of $A_{\alpha }$ nucleons. As we deal with two body
potential, then we can perform integration over all single-particle
coordinates $\mathbf{r}_{i}^{\prime }$ ($\mathbf{r}_{j}^{\prime }$) but one.
As a results integration over all but one coordinates leads us to the
density distribution
\begin{equation}
  \rho _{\alpha }\left( \mathbf{r}\right) =
  \displaystyle\int dV_{\alpha }^{\prime }\Phi_{\alpha }\left( A_{\alpha }\right)
  \displaystyle\sum_{i}\delta \left( \mathbf{r-r}_{i}^{\prime }\right)
  \Phi _{\alpha }\left( A_{\alpha }\right).
\label{eq:F07}
\end{equation}
And thus
\begin{equation}
  \widehat{V}^{(F)}(\mathbf{r}) =
  \displaystyle\int d\mathbf{r}_{1}\, d\mathbf{r}_{2}\,
  \rho_{1} \left( \mathbf{r}_{1}\right) \widehat{V}(\mathbf{r}_{1}-\mathbf{r}_{2} + \mathbf{r})\,
  \rho_{2} \left( \mathbf{r}_{2}\right).
\label{eq:F08}
\end{equation}
For $s$-nuclei density distribution equals
\begin{equation}
  \rho _{\alpha }\left( \mathbf{r}_{\alpha }\right) = N_{\alpha }\exp \left\{ -%
  \displaystyle\frac{r^{2}}{b^{2}}\, \displaystyle\frac{A_{\alpha }}{A_{\alpha }-1}\right\},
\label{eq:F09}
\end{equation}
where $b$ is the oscillator length.

By using Fourier transformation we can reduce (\ref{eq:F08}) to the form
\begin{equation}
  \widehat{V}^{(F)}(\mathbf{r}) =
  \int d\mathbf{k}\; \mathcal{V}(\mathbf{k})\:
  \exp \{i\, \mathbf{k}\, \mathbf{r}
  \}\, \Phi _{1}(\mathbf{k})\, \Phi _{2}(\mathbf{k}).
\label{eq:F09}
\end{equation}
Here, $\mathcal{V}$ denotes the Fourier transform of a nucleon-nucleon
interaction
\begin{equation}
  \widehat{V}(\mathbf{r}) =
  (2\pi )^{-3/2} \displaystyle\int d\mathbf{k}\, \exp \{-i\mathbf{k} \mathbf{r}\}\, \mathcal{V}(\mathbf{k})
\end{equation}
and $\Phi _{\alpha }(\mathbf{k})$ denotes form-factor of subsystem with $%
A_{\alpha }$ nucleons. For the $s$-shell nuclei, it easy to find that
\begin{equation}
  \Phi_{\alpha}(\mathbf{k}) =
  \exp \left\{ - \displaystyle\frac{k^{2}b^{2}}{4}\,
  \displaystyle\frac{A_{\alpha} - 1}{A_{\alpha }}\right\}.
\end{equation}
For a nucleon-nucleon interaction, having Gaussian form
\begin{equation}
  \widehat{V}(ij) =
  V_{0}\,\exp \left\{ - \displaystyle\frac{(\mathbf{r}_{i} - \mathbf{r}_{j})^{2}}{a^{2}}\right\}
\end{equation}
all calculation can be done analytically in closed form and results reads:
\begin{equation}
  \widehat{V}^{(F)}(\mathbf{r}) =
  V_{0}\,z^{3/2}\exp \left\{ - \displaystyle\frac{r^{2}}{a^{2}}z\right\},
\end{equation}
where
\begin{equation}
\begin{array}{llllll}
  z = \left( 1 + \displaystyle\frac{b^{2}}{a^{2}}[2-\mu ^{-1}]\right) ^{-1}.
\end{array}
\end{equation}

It is worth while to notice, that when ratio $b^{2}/a^{2}$ is rather small
(it takes place for a very wide potential well), then $z\approx 1$ and the
folding potential almost coincides with the nucleon-nucleon potential. In
other limit case, when $b^{2}/a^{2}\gg 1$ (it may realize for potential with
small radius of a core), then intensity, as well as radius of potential are
significantly redetermined.

\subsection{Case of two-cluster systems
\label{sec.folding.twoclusters}}

General formulas, obtained above, may be adopted to the case of interest,
namely, to two-cluster systems. Here we consider two-cluster systems where
one of the clusters is an alpha-particle ($A_{1}$=4) and the second cluster
consists of $A_{2}$ nucleons with \ 1$\leq A_{2}\leq $4. In this case we can
write :
\begin{equation}
\begin{array}{llllll}
  V_{NN}^{(F)} & = &
  \displaystyle\frac{A_{2}}{4}(9V_{33}+3V_{31}+3V_{13}+V_{11}) 
  \left( 1 + \displaystyle\frac{b^{2}}{a^{2}} \displaystyle\frac{3}{4}\right) ^{-3/2}\exp
  \left\{ -\, \displaystyle\frac{R^{2}}{a^{2}}
    \Bigl( 1 +  \displaystyle\frac{b^{2}}{a^{2}}  \displaystyle\frac{3}{4} \Bigr)^{-1}
  \right\},
\end{array}
\label{eq:F11}
\end{equation}
\begin{equation}
V_{NN}^{(F)}(\mathbf{r}) = \frac{A_{2}}{4}(9V_{33}+3V_{31}+3V_{13}+V_{11})
 z^{3/2}\exp \left\{ -\frac{R^{2}}{a^{2}}z\right\} ,
 \label{eq:F11A}
\end{equation}
where
\begin{eqnarray}
z &=&\left( 1+\frac{b^{2}}{a^{2}}\left[ 2-\mu ^{-1}\right] \right) ^{-1}, \nonumber\\
\mu  &=&\frac{A_{1}A_{2}}{A_{1}+A_{2}} \nonumber
\end{eqnarray}
and where $V_{33}$, $V_{31}$, $V_{13}$, $V_{11}$ are intensities of the central
nucleon-nucleon interaction (denoted as $V_{2S+1,2T+1}$) with the fixed
value of the spin $S$ and isospin $T$ of interaction nucleons. Each
component of the potential is presented by sum of two or three Gaussians
\begin{equation}
\begin{array}{llllll}
  V_{2S+1,2T+1} \left( r\right) =
  \displaystyle\sum_{i=1}^{N_{G}} V_{2S+1,2T+1}^{\left( i\right) }
  \exp \left\{ - \displaystyle\frac{r^{2}}{a_{i}^{2}}\right\}.
\end{array}
\end{equation}
Expressions, obtained for $NN$-interaction with Gauss spatial form, can be
easily transformed to the case of the Coulomb forces. For this aim we shall
use the well-known relation:
\begin{equation}
  \displaystyle\frac{1}{r} = \displaystyle\frac{2}{\sqrt{\pi }} \displaystyle\int_{0}^{\infty}\, dx\, \exp \{-r^{2}x^{2}\}.
\end{equation}
Then Coulomb interaction between cluster with the number of protons $Z_{1}$ and $Z_{2}$
\begin{equation}
  \widehat{V}_{C}^{\left( F\right) } (\mathbf{r}) =
  \displaystyle\frac{Z_{1}\, Z_{2}\, e^{2}}{b}
  \displaystyle\frac{2}{\sqrt{\pi}}\, \displaystyle\int d\gamma \,z^{3/2}\exp \left\{ -\, \displaystyle\frac{r^{2}}{b^{2}}\gamma
^{2}z\right\} .  \label{eq:F13A}
\end{equation}
By introducing new variable for integration:
\begin{equation}
  a = \displaystyle\frac{\sigma \gamma ^{2}}{1 + \sigma \gamma ^{2}},\,\sigma =
  2 - \mu ^{-1},
\end{equation}
we obtain the integral
\begin{equation}
  \widehat{V}_{C}^{\left( F\right) } (\mathbf{r}) =
  \displaystyle\frac{Z_{1}Z_{2}e^{2}}{b}\frac{2}{\sqrt{\pi }}
  \displaystyle\frac{1}{2\sqrt{\sigma }}\,
  \displaystyle\int_{0}^{1} da\, a^{-1/2}\,
  \exp \left\{ -\,\displaystyle\frac{r^{2}}{\sigma \,b^{2}}a\right\},
\label{eq:F13B}
\end{equation}
which leads to the error function:
\begin{equation}
  \widehat{V}_{C}^{\left( F\right) } (\mathbf{r}) =
  \displaystyle\frac{Z_{1}\, Z_{2}\, e^{2}}{R}\;
  {\rm erf}\left( \displaystyle\frac{r^{2}}{\sigma \,b^{2}}\right).
\label{eq:F13C}
\end{equation}
For large values of $R\gg 1$, we have got
\begin{equation}
  \widehat{V}_{C}^{\left( F\right) } (\mathbf{r}) \approx \displaystyle\frac{Z_{1}\, Z_{2}\, e^{2}}{r}.
\end{equation}

\subsection{Operator of emission of bremsstrahlung photons
\label{sec.operatoremission}}

The translation invariant operator of interaction of photon with atomic nuclei is%
\begin{equation}
\widehat{H}_{e}\left( \mathbf{k}_{\gamma },\mathbf{\varepsilon }_{\mu
}\right) =\frac{1}{2}\frac{e\hbar }{m_{N}c}\sum_{i=1}^{A}\frac{1}{2}\left( 1+%
\widehat{\tau }_{iz}\right) \left[ {\hat{\pibf}}_{i}^{\ast }%
\mathbf{A}^{\ast }\left( i\right) +\mathbf{A}^{\ast }\left( i\right)
\widehat{\pibf}_{i}^{\ast }\right]
\label{eq.operatoremission.1}
\end{equation}
where
\begin{equation}
\begin{array}{lllllllll}
\vspace{1.5mm}
  \mathbf{A}^{\ast }\left( i\right) =
  \mathbf{\varepsilon }_{\mu }\exp
  \left\{ -i\left( \mathbf{k}_{\gamma} \rhobf_{i}\right) \right\}, &
  \widehat{\pibf}_{i}^{\ast } = i\nabla _{\rhobf_{i}}, &
  \rhobf_{i} = \mathbf{r}_{i}-\mathbf{R}_{cm}, \\

  \mathbf{R}_{cm} = \displaystyle\frac{1}{A} \displaystyle\sum_{i=1}^{A}\mathbf{r}_{i}, &
  \widehat{\pibf}_{i} = \widehat{\mathbf{p}}_{i}-\widehat{\mathbf{P}}_{cm}, &
  \widehat{\mathbf{P}}_{cm} = \displaystyle\frac{1}{A} \displaystyle\sum_{i=1}^{A}\widehat{\mathbf{p}}_{i},
\end{array}
\label{eq.operatoremission.2}
\end{equation}
%
and $\mathbf{k}_{\gamma }$\ is wave vector of photon and $\mathbf{%
\varepsilon }_{\mu }$ is its circular polarization.

It is worthwhile noticing that operator $\widehat{H}_{e}\left( \mathbf{k}%
_{\gamma },\mathbf{\varepsilon }_{\mu }\right) $ has to be projected
momentum $\lambda $ which determine the multipolity of the emitted photon.
The projected operator we denote as $\widehat{H}_{e}^{\left( \lambda \right)
}\left( \mathbf{k}_{\gamma },\mathbf{\varepsilon }_{\mu }\right) $. The
projected operator $\widehat{H}_{e}^{\left( \lambda \right) }\left( \mathbf{k%
}_{\gamma },\mathbf{\varepsilon }_{\mu }\right) $ will be proportional to
spherical functions $Y_{\lambda \mu }\left( \widehat{\mathbf{r}}_{i}\right) $%
. We do not dwell on projection of the operator $\widehat{H}_{e}\left(
\mathbf{k}_{\gamma },\mathbf{\varepsilon }_{\mu }\right) $, we will project
matrix elements of this operator calculated between Slater determinants.
Later we will also consider the operator
\begin{equation}
\begin{array}{lllllllll}
  \widehat{H}_{0} =
  \displaystyle\sum_{i=1}^{A} \displaystyle\frac{1}{2}
  \left( 1 + \widehat{\tau }_{iz}\right)
  \exp \left\{ i\left( \mathbf{k}_{\gamma }\mathbf{r}_{i}\right)
\right\} ,
\end{array}
\label{eq.operatoremission.3}
\end{equation}
%
which determines emission or absorption of photon in the capture reaction or
photodisintegration reaction, respectively. This operator can be easily
projected on quantum number $\lambda$
\begin{equation}
\begin{array}{lllllllll}
  \widehat{H}_{0}^{\left( \lambda \right) } =
  \displaystyle\sum_{i=1}^{A} \displaystyle\frac{1}{2}\left( 1 + \widehat{\tau }_{iz}\right) j_{\lambda }
  \left( k_{\gamma }r_{i}\right) Y_{\lambda \mu }\left( \widehat{\mathbf{r}}_{i}\right)
\end{array}
\label{eq.operatoremission.4}
\end{equation}

To calculate matrix elements of the operator between Slater determinant wave
functions, it is more expedient to use the single-particle operator%
\begin{equation}
\widehat{\overline{H}}_{e}\left( \mathbf{k}_{\gamma },\mathbf{\varepsilon }%
_{\mu }\right) =-\frac{1}{2}\frac{e\hbar }{m_{N}c}\sum_{i=1}^{A}\frac{1}{2}%
\left( 1+\widehat{\tau }_{iz}\right) \left[ \mathbf{\varepsilon }_{\mu }\exp
\left\{ -i\left( \mathbf{k}_{\gamma }\mathbf{r}_{i}\right) \right\} \widehat{%
\mathbf{p}}_{i}^{\ast }\right].
\label{eq.operatoremission.5}
\end{equation}
After calculations we obtain
%
\begin{equation}
\begin{array}{lllllll}
\vspace{0.5mm}
  \widehat{\overline{H}}_{e}\left( \mathbf{k}_{\gamma },\mathbf{\varepsilon}_{\mu }\right)

  & = &
  -\displaystyle\frac{1}{2}\frac{e\hbar }{m_{N}c}
  \displaystyle\sum_{i=1}^{A}\frac{1}{2}\left( 1+%
  \widehat{\tau }_{iz}\right) \left[ \mathbf{\varepsilon }_{\mu }\exp \left\{
  -i\left( \mathbf{k}_{\gamma },\mathbf{r}_{i}-\mathbf{R}_{cm}\right) \right\}
  \left( \widehat{\mathbf{p}}_{i}^{\ast }-\widehat{\mathbf{P}}_{cm}\right) %
  \right]

  \times \exp \left\{ -i\left( \mathbf{k}_{\gamma },\mathbf{R}_{cm}\right) \right\} \\

  & - &
  \displaystyle\frac{1}{2}\frac{e\hbar }{m_{N}c}
  \displaystyle\sum_{i=1}^{A}\frac{1}{2}\left( 1+%
  \widehat{\tau }_{iz}\right) \exp \left\{ -i\left( \mathbf{k}_{\gamma },%
  \mathbf{r}_{i}-\mathbf{R}_{cm}\right) \right\} 
  \times
    \exp \left\{ -i\left( \mathbf{k}_{\gamma },\mathbf{R}_{cm}\right)
\right\} \left( \mathbf{\varepsilon }_{\mu }\widehat{\mathbf{P}}_{cm}\right).
\end{array}
\label{eq.operatoremission.6}
\end{equation}

\subsection{Cross section of bremsstrahlung emission
\label{sec.crosssection}}

We follow papers~\cite{1990PhRvC..41.1401L,1990PhRvC..42.1895L} of Liu, Tang and Kanada
to consider the bremsstrahlung emission in light nuclei.
The differential cross section of the bremsstrahlung emission in the complanar laboratory framework is
\begin{equation}
\begin{array}{lllll}
\vspace{1.0mm}
  \displaystyle\frac{d\, \sigma^{(1)}}{d\Omega_{1}\, d\Omega_{2}\, d\Omega_{\gamma}} & = &
  \displaystyle\frac{E_{\gamma}}{(2\pi\hbar)^{4}}\,
  \Bigl( \displaystyle\frac{p_{f}}{\hbar c} \Bigr)\,
    \displaystyle\frac{\sin^{2} \theta_{1}\, \sin^{2} \theta_{2} }{\sin^{5}  (\theta_{1} + \theta_{2})}\; 

  \displaystyle\frac{1}{2J + 1}
  \displaystyle\sum\limits_{\mu m_{i}}
    \Bigl|
      \Bigl\langle \Psi_{E_{f} l_{f}} \Bigl| \hat{H}_{\gamma} (\vb{k}_{\gamma}, \varepsilon_{\mu}) \Bigr| \Psi_{E_{i}l_{i}} \Bigr\rangle
    \Bigr|^{2}.
\end{array}
\label{eq.crosssection.1.1}
\end{equation}
New version
\begin{equation}
  \frac{d^{3}\sigma^{(2)} }{d\Omega_{1} d\Omega _{2} dE_{\gamma }} =
  \frac{p_{1}^{4}v_{f}}{\left( 2\pi \hbar \right) ^{4}\hbar }
  \frac{\sin ^{2}\theta_{1}\sin ^{2}\theta _{2}}{\sin ^{5}\left( \theta _{1}+\theta _{2}\right) }
  \sum_{\mu }\left\vert \left\langle \Psi_{E_{f} l_{f}}
  \left\vert \widehat{H}_{e}\left(
  \mathbf{k}_{\gamma },\mathbf{\varepsilon }_{\mu }\right) \right\vert \Psi_{E_{i}l_{i}}
  \right\rangle \right\vert^{2},
\label{eq.crosssection.1.2}
\end{equation}
where $p_{1}$ is the momentum of the incident nucleus (cluster) with $A_{1}$ nucleons.

The kinematic relations for initial ($E_{i}$), final  ($E_{f}$) energies of
a two-cluster system and the photon energy ($E_{\gamma }$) are
\begin{equation}
E_{i}=E_{f}+E_{\gamma }
\label{eq.crosssection.1.3}
\end{equation}
and
\begin{equation}
  E_{\gamma } =
  E_{1,i}\left[ 1-\frac{1}{A_{2}}\frac{A_{1}\sin ^{2}\theta
_{1}+A_{2}\sin ^{2}\theta _{2}}{\sin ^{2}\left( \theta _{1}+\theta
_{2}\right) }\right],
\label{eq.crosssection.1.4}
\end{equation}
where $E_{1,i}$ is the energy of the incident cluster $A_{1}$. Energies ($%
E_{i}$) and ($E_{f}$)  are determined in the center of mass of coordinate
system.

\section{Matrix elements in folding approximation
\label{sec.folding.1}}

Matrix element of bremsstrahlung emission of photons for two $s$-clusters
(i.e., for clusters with $1 \le A_{\alpha} \le 4$ or
for $n$, $p$, $d$, $^{3}{\rm H}$,
$^{3}{\rm He}$, $^{4}{\rm He}$) is
[see App.~\ref{sec.app.folding} for details]
%
\begin{equation}
\begin{array}{lllll}
\vspace{1.0mm}
  & \Bigl\langle \Psi_{E_{f} l_{f}} \Bigl| \hat{H}_{\gamma} (\vb{k}_{\gamma}, \varepsilon_{\mu}) \Bigr| \Psi_{E_{i} l_{i}} \Bigr\rangle = \\

\vspace{1.0mm}
  = &
  \sqrt{\displaystyle\frac{A_{2}}{A_{1}\, A}}
  \biggl\langle
    R_{E_{f} l_{f}} (r)\, Y_{l_{f} m_{f}} (\widehat{\mathbf{r}}_{i})
    \biggl|
      \exp{-i\, \sqrt{\displaystyle\frac{A_{2}}{A_{1}\, A}}\, (\vb{k}_{\gamma}, \mathbf{r})}\,
      (\varepsilonbf_{\mu}, \hat{\pibf})
    \biggr|
    R_{E_{i} l_{i}} (r)\, Y_{l_{i} m_{i}} (\widehat{\mathbf{r}}_{i})
  \biggr\rangle\, F_{1} - \\

  - &
  \sqrt{\displaystyle\frac{A_{1}}{A_{2}\, A}}
  \biggl\langle
    R_{E_{f} l_{f}} (r)\, Y_{l_{f} m_{f}} (\widehat{\mathbf{r}}_{i})
    \biggl|
      \exp{i\, \sqrt{\displaystyle\frac{A_{1}}{A_{2}\, A}}\, (\vb{k}_{\gamma}, \vb{r})}\,
      (\varepsilonbf_{\mu}, \hat{\pibf})
    \biggr|
    R_{E_{i} l_{i}} (r)\, Y_{l_{i} m_{i}} (\widehat{\mathbf{r}}_{i})
  \biggr\rangle\, F_{2}.
\end{array}
\label{eq.folding.2.1}
\end{equation}
In the standard approximation of resonanting group method, form factor $F_{\alpha}$ equals ($\alpha=1,2$)
\begin{equation}
\begin{array}{lllll}
  F_{\alpha} = &
  \Bigl\langle \Phi_{\alpha} (A_{\alpha}) \Bigl| F_{0}^{(\alpha)} \Bigr| \Phi_{\alpha}  (A_{\alpha}) \Bigr\rangle =
  Z_{\alpha}\, \exp{ - \displaystyle\frac{1}{4}\, \displaystyle\frac{A_{\alpha} - 1}{A_{\alpha}}\, (k, b)^{2}},
\end{array}
\label{eq.folding.2.2}
\end{equation}
with $b$ is oscillator length. Thus, to determine cross section of the
bremsstrahlung emission, we need to calculate matrix element ($i = 1,2$)
\begin{equation}
\begin{array}{lllll}
\vspace{2.0mm}
  \mathbf{I}\, (\alpha_{i}) =
  \left\langle
    R_{E_{f} l_{f}} (r)\, Y_{l_{f} m_{f}} (\widehat{\mathbf{r}})
    \Bigl| \exp \left\{ - i\alpha_{i}
      \left( \mathbf{k}_{\gamma },\mathbf{r}\right) \right\}
      \widehat{\pibf}
    \Bigr|
    R_{E_{i} l_{i}} (r)\,
    Y_{l_{i} m_{i}} (\widehat{\mathbf{r}})
  \right\rangle, \\

  I_{\mu}\, (\alpha_{i}) =
  \mathbf{\varepsilonbf}_{\mu}\, \mathbf{I}\, (\alpha_{i}) =
  \left\langle
    R_{E_{f} l_{f}} \left( r \right) Y_{l_{f} m_{f}} \left( \widehat{\mathbf{r}} \right)
    \Bigl| \exp \left\{- i\alpha_{i}
      \left( \mathbf{k}_{\gamma },\mathbf{r}\right) \right\}
      \left( \mathbf{\varepsilonbf}_{\mu }\widehat{\pibf} \right)
    \Bigr|
    R_{E_{i} l_{i}} (r)\,
    Y_{l_{i} m_{i}} (\widehat{\mathbf{r}})
  \right\rangle
\end{array}
\label{eq.folding.2.3}
\end{equation}
for two values of the parameter
\begin{equation}
  \alpha_{1} = \sqrt{\frac{A_{2}}{A_{1}A}}, \quad
  \alpha_{2} = - \sqrt{\frac{A_{1}}{A_{2}A}}.
\label{eq.folding.2.4}
\end{equation}

\subsection{Multipole expansion
\label{sec.multiple}}

Applying the multipolar expansion, the integral is
[see App.~\ref{sec.app.multipole}, Eq.~(\ref{eq.2.4.4.1})] 
\begin{equation}
\begin{array}{ll}
  \vb{I}_{1} (\alpha_{i}) =
  \biggl< \phi_{f} \biggl| \,  e^{-i\alpha_{i} \mathbf{k_{\gamma}r}}
  \displaystyle\frac{\partial}{\partial \vb{r}}\, \biggr| \, \phi_{i} \biggr>_\mathbf{r} =
  \sqrt{\displaystyle\frac{\pi}{2}}\:
  \displaystyle\sum\limits_{l_{\gamma}=1}\,
    (-i)^{l_{\gamma}}\, \sqrt{2l_{\gamma}+1}\;
  \displaystyle\sum\limits_{\mu = \pm 1}
    \xibf_{\mu}\, \mu\,
    \Bigl[ p_{l_{\gamma}\mu}^{M} - i\mu\: p_{l_{\gamma}\mu}^{E} \Bigr],
\end{array}
\label{eq.multiple.1}
\end{equation}
where (at $l_{i} = 0$)
\begin{equation}
\begin{array}{lcl}
\vspace{3mm}
  p_{l_{\gamma}\mu}^{M 0 m_{f}} 
   & = &
    - I_{M}(0, l_{f}, l_{\gamma}, 1, \mu) \cdot J_{1}(0, l_{f},l_{\gamma}, \alpha_{i}), \\

\vspace{1mm}
  p_{l_{\gamma}\mu}^{E 0 m_{f}} & = &
    \sqrt{\displaystyle\frac{l_{\gamma}+1}{2l_{\gamma}+1}} \cdot I_{E} (0,l_{f},l_{\gamma}, 1, l_{\gamma}-1, \mu) \cdot J_{1}(0,l_{f},l_{\gamma}-1, \alpha_{i})\; - \\
  & - &
    \sqrt{\displaystyle\frac{l_{\gamma}}{2l_{\gamma}+1}} \cdot I_{E} (0,l_{f}, l_{\gamma}, 1, l_{\gamma}+1, \mu) \cdot J_{1}(0,l_{f},l_{\gamma}+1, \alpha_{i})
\end{array}
\label{eq.multiple.2}
\end{equation}
and
\begin{equation}
\begin{array}{ccl}
  J_{1}(l_{i},l_{f},n, \alpha_{i}) & = &
  \displaystyle\int\limits^{+\infty}_{0}
     \displaystyle\frac{dR_{i}(l_{i}, r)}{dr}\: R^{*}_{f}(l_{f},r)\, j_{n}(\alpha_{i} kr)\; r^{2} dr.
\end{array}
\label{eq.multiple.3}
\end{equation}
Here, $\varepsilonbf^{(\alpha)}$ are unit vectors of \emph{linear} polarization of the photon emitted ($\varepsilonbf^{(\alpha), *} = \varepsilonbf^{(\alpha)}$), $\mathbf{k}_{\gamma}$ is wave vector of the photon and $w_{\gamma} = k_{\gamma} c = \bigl| \mathbf{k}_{\gamma}\bigr|c$. Vectors $\varepsilonbf^{(\alpha)}$ are perpendicular to $\mathbf{k}_{\gamma}$ in Coulomb gauge. We have two independent polarizations $\varepsilonbf^{(1)}$ and $\varepsilonbf^{(2)}$ for the photon with impulse $\mathbf{k}_{\gamma}$ ($\alpha=1,2$).
$\xibf_{\mu}$ are \emph{vectors of circular polarization} with opposite directions of rotation (see Ref.~\cite{Eisenberg.1973}, (2.39), p.~42;
see App.~\ref{sec.app.polarization}).
Also we have properties:
\begin{equation}
\begin{array}{lclc}
  \Bigl[ \mathbf{k}_{\gamma} \times \varepsilonbf^{(1)} \Bigr] = k_{\gamma}\, \varepsilonbf^{(2)}, &
  \Bigl[ \mathbf{k}_{\gamma} \times \varepsilonbf^{(2)} \Bigr] = -\, k_{\gamma}\, \varepsilonbf^{(1)}, &
  \Bigl[ \mathbf{k}_{\gamma} \times \varepsilonbf^{(3)} \Bigr] = 0, &
%
  \displaystyle\sum\limits_{\alpha=1,2,3}
  \Bigl[ \mathbf{k}_{\gamma} \times \varepsilonbf^{(\alpha)} \Bigr] =
  k_{\gamma}\, (\varepsilonbf^{(2)} - \varepsilonbf^{(1)}).
\end{array}
\label{eq.multiple.4}
\end{equation}
%
Also we have property:
\begin{equation}
\begin{array}{llll}
  \displaystyle\sum\limits_{\alpha=1,2} \varepsilonbf^{(\alpha)} \cdot \vb{I}_{1} (\alpha_{i}) =
  \sqrt{\displaystyle\frac{\pi}{2}}\:
  \displaystyle\sum\limits_{l_{\gamma}=1}\, (-i)^{l_{\gamma}}\, \sqrt{2l_{\gamma}+1}\;
  \displaystyle\sum\limits_{\mu=\pm 1} \mu\,h_{\mu}\, \bigl(p_{l_{\gamma}, \mu}^{M} + p_{l_{\gamma}, -\mu}^{E} \bigr), \\

\end{array}
\label{eq.multiple.5}
\end{equation}
where
[see Eqs.~(\ref{eq.app.polarization.5}), App.~\ref{sec.app.polarization}]
\begin{equation}
\begin{array}{llllll}
  h_{\pm} = \mp \displaystyle\frac{1 \pm i}{\sqrt{2}},
  \hspace{10mm}
  h_{-} + h_{+} = -i\, \sqrt{2}, &

  \hspace{10mm}
  \displaystyle\sum\limits_{\mu=\pm 1} \mu\,h_{\mu} =
  - h_{-} + h_{+} =
  - \sqrt{2}.
\end{array}
\label{eq.multiple.6}
\end{equation}


\subsection{Case of $l_{i}=0$, $l_{f}=1$, $l_{\gamma}=1$
\label{sec.simplecase}}

In a case of $l_{i}=0$, $l_{f}=1$, $l_{\gamma}=1$ integral (\ref{eq.multiple.1}) is simplified to
\begin{equation}
\begin{array}{llllll}
  \vb{I}_{1} (\alpha_{i}) =
  -i\, \sqrt{\displaystyle\frac{3\pi}{2}}\:
  \displaystyle\sum\limits_{\mu = \pm 1}
    \xibf_{\mu}\, \mu\;
    \Bigl[ p_{l_{\gamma}\mu}^{M} - i\mu\: p_{l_{\gamma}\mu}^{E} \Bigr],
\end{array}
\label{eq.simplecase.1.1}
\end{equation}
where [see Eqs.~(\ref{eq.multiple.2})]
\begin{equation}
\begin{array}{lll}
\vspace{3mm}
  p_{l_{\gamma}\mu}^{M 0 m_{f}} = - I_{M}(0, 1, 1, 1, \mu) \cdot J_{1}(0, 1, 1, \alpha_{i}), \\

  p_{l_{\gamma}\mu}^{E 0 m_{f}} =
    \sqrt{\displaystyle\frac{2}{3}}\, I_{E} (0,1,1,1,0, \mu) \cdot J_{1}(0,1,0, \alpha_{i}) -
    \sqrt{\displaystyle\frac{1}{3}}\, I_{E} (0,1,1,1,2, \mu) \cdot J_{1}(0,1,2, \alpha_{i}).
\end{array}
\label{eq.simplecase.1.2}
\end{equation}
Results of calculation of angular integrals are
[we omit details of calculations in this paper]:
\begin{equation}
\begin{array}{llllll}
\vspace{0.8mm}
  I_{E}\, (0, 1, 1, 1, 0, \mu) = \sqrt{\displaystyle\frac{1}{24\pi}}, &
  I_{M}\, (0, 1, 1, 1, \mu) = 0, &
  I_{E}\, (0, 1, 1, 1, 2, \mu) = \displaystyle\frac{47}{240} \sqrt{\displaystyle\frac{3}{2\pi}},
\end{array}
\label{eq.simplecase.1.3}
\end{equation}
and matrix elements (\ref{eq.simplecase.1.2}) are simplified to
\begin{equation}
\begin{array}{lllllllll}
\vspace{2mm}
  p_{l_{\gamma}\mu}^{M 0 m_{f}} = 0, &
  p_{l_{\gamma}\mu}^{E 0 m_{f}} =
    \displaystyle\frac{1}{6} \sqrt{\displaystyle\frac{1}{\pi}} \cdot J_{1}(0,1,0, \alpha_{i}) -
    \displaystyle\frac{47}{240} \sqrt{\displaystyle\frac{1}{2\pi}} \cdot J_{1}(0,1,2, \alpha_{i}).
\end{array}
\label{eq.simplecase.1.4}
\end{equation}
We substitute these solutions to Eq.~(\ref{eq.simplecase.1.1}) and obtain:
\begin{equation}
\begin{array}{llllll}
  \vb{I}_{1} (\alpha_{i}) & = &
  - \displaystyle\frac{1}{6}\;
  \sqrt{\displaystyle\frac{3}{2}}\:
  \Bigl(
    J_{1}(0,1,0, \alpha_{i}) -
    \displaystyle\frac{47}{40} \sqrt{\displaystyle\frac{1}{2}} \cdot J_{1}(0,1,2, \alpha_{i})
  \Bigr)\;
  \bigl( \xibf_{\mu=+1} + \xibf_{\mu=-1} \bigr).
\end{array}
\label{eq.simplecase.1.7}
\end{equation}

\subsection{Action on vectors of polarization
\label{sec.polarize}}

Now we calculate summation over vectors of polarization.
We use definition of vectors of polarization [see Eq.~(\ref{eq.app.polarization.1})],
and find:
\begin{equation}
\begin{array}{cc}
  \varepsilonbf^{(1)} = \displaystyle\frac{1}{\sqrt{2}}\, \bigl(\xibf_{-1} - \xibf_{+1}\bigr), &
  \varepsilonbf^{(2)} = \displaystyle\frac{i}{\sqrt{2}}\, \bigl(\xibf_{-1} + \xibf_{+1}\bigr).
\end{array}
\label{eq.polirize.2}
\end{equation}
Using (\ref{eq.simplecase.1.7}), we obtain:
\begin{equation}
\begin{array}{llll}
  \varepsilonbf^{(1)} \cdot \bigl( \xibf_{\mu=+1} + \xibf_{\mu=-1} \bigr) = 0, &
  \varepsilonbf^{(2)} \cdot \bigl( \xibf_{\mu=+1} + \xibf_{\mu=-1} \bigr) = i\, \sqrt{2}.
\end{array}
\label{eq.polirize.4}
\end{equation}
On such a basis, from Eq.~(\ref{eq.simplecase.1.7}) we find:
\begin{equation}
\begin{array}{llllll}
\vspace{1.5mm}
  \varepsilonbf^{(1)} \cdot \vb{I}_{1} (\alpha_{i}) & = & 0, &

\vspace{1.0mm}
  \varepsilonbf^{(2)} \cdot \vb{I}_{1} (\alpha_{i}) & = &
  -\, i\;
  \displaystyle\frac{\sqrt{3}}{6}\,
  \Bigl(
    J_{1}(0,1,0, \alpha_{i}) -
    \displaystyle\frac{47}{40} \sqrt{\displaystyle\frac{1}{2}}\, J_{1}(0,1,2, \alpha_{i})
  \Bigr).
\end{array}
\label{eq.polirize.5}
\end{equation}
Using definition (\ref{eq.multiple.1}) for $\vb{I_{1}}\, (\pm\alpha)$,
the matrix element (\ref{eq.folding.2.1}) is written down as
\begin{equation}
\begin{array}{lllll}
  \Bigl\langle \Psi_{E_{f} l_{f}} \Bigl| \hat{H}_{\gamma} (\vb{k}_{\gamma}, \varepsilonbf_{\mu}) \Bigr| \Psi_{E_{i}l_{i}} \Bigr\rangle =
  -\, i \hbar\,
  \Bigl\{
  \sqrt{\displaystyle\frac{A_{2}}{A_{1}\, A}}\,
    \varepsilonbf_{\mu}\, \vb{I}_{1} \Bigl( +\, \sqrt{\displaystyle\frac{A_{2}}{A_{1}\, A}}\, \Bigr)\, F_{1} -

  \sqrt{\displaystyle\frac{A_{1}}{A_{2}\, A}}\,
    \varepsilonbf_{\mu}\, \vb{I}_{1} \Bigl( -\, \sqrt{\displaystyle\frac{A_{1}}{A_{2}\, A}}\, \Bigr)\, F_{2}
  \Bigr\}.
\end{array}
\label{eq.simplestcase.1.9}
\end{equation}

\section{Analysis
\label{sec.analysis}}

Cross sections of bremsstrahlung emission are determined on the basis of integrals defined in Eqs.~(\ref{eq.multiple.3})
[see Eqs.~(\ref{eq.crosssection.1.2}) or (\ref{eq.crosssection.1.3}), Eqs.~(\ref{eq.simplestcase.1.9}), (\ref{eq.polirize.5})].
These integrals involve radial wave functions $R_{El}(r)$.
These radial wave functions are calculated numerically by solving the Schr\"{o}dinger equation with the corresponding potential of interaction between two studied clusters (nuclei).
Some details of normalization, asymptotic behavior of such wave functions and their numeric calculations are given in Appendix~\ref{sec.app.4}.
In Fig.~\ref{fig.potentials.1} such cluster-cluster potentials constructed within our formalism above are presented.
These potentials include nuclear and Coulomb components and are determined by using the Minnesota nucleon-nucleon potential
\cite{kn:Minn_pot1, 1970NuPhA.158..529R}
and formulas (\ref{eq:F11A}) and (\ref{eq:F13C}).
%
\begin{figure}[htbp]
\centerline{\includegraphics[width=88mm]{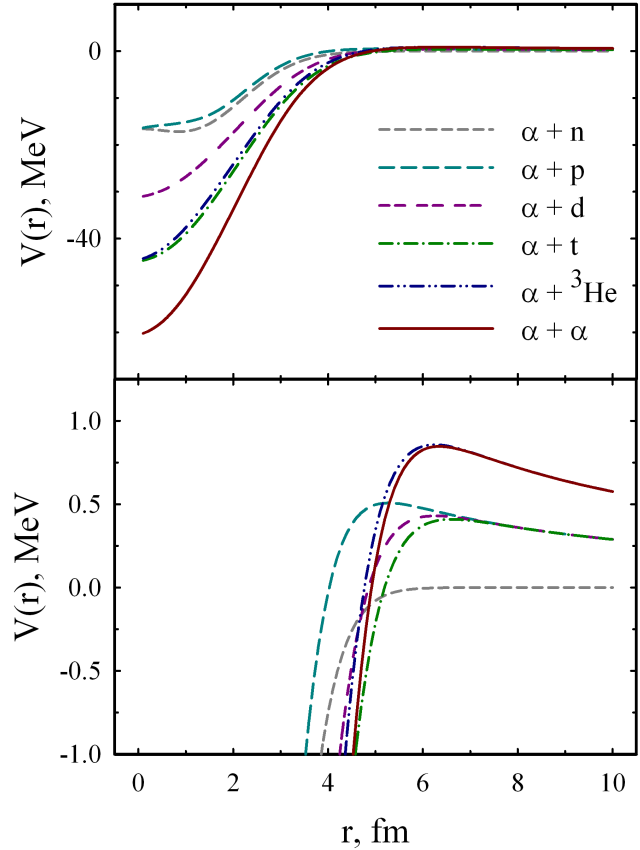}}
\vspace{-2mm}
\caption{\small (Color online)
Potentials for different nuclear systems
[see text in Sect.~\ref{sec.analysis} for details].
\label{fig.potentials.1}}
\end{figure}
Also in this paper we will restrict ourselves by the bremsstrahlung cross sections integrated over angles.

\subsection{Which parameters have essential influence on the accuracy of calculations of the spectra
\label{sec.crosssection.2}}

Achieving the needed accuracy of calculation of the spectra is important in this problem.
It turns out that just direct derivation of the bremsstrahlung spectra at some chosen parameters of calculations does not give satisfactory convergence in calculations for some nuclei.
It causes necessity to understand what determines the accuracy of the calculations and what causes errors in the calculation on the computer.
Note that it would seem that it would be possible to increase the number of intervals in the selected region of integration in order to obtain higher convergent calculations of the spectra.
By such a way, we chose minimum number of intervals equal to 4 per one oscillation of the radial wave function.
However, it turned out that just increasing the number of intervals did not allow us to increase the accuracy of determining the cross section.
From formalism one can find that it is more important to analyze the full integrant function of the radial integral, rather than the wave functions themselves.

\begin{figure}[htbp]
\centerline{\includegraphics[width=88mm]{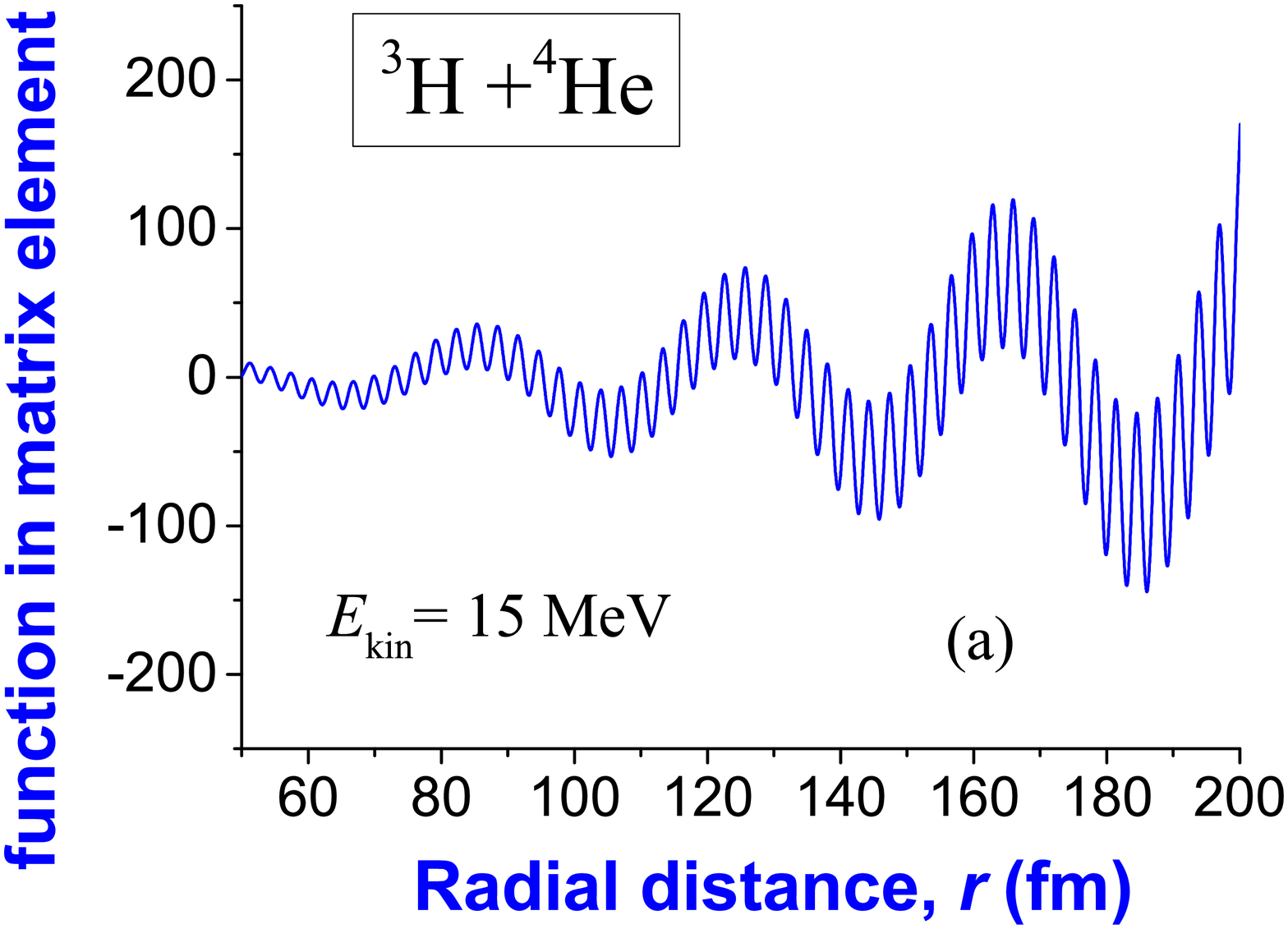}
\hspace{-5mm}\includegraphics[width=88mm]{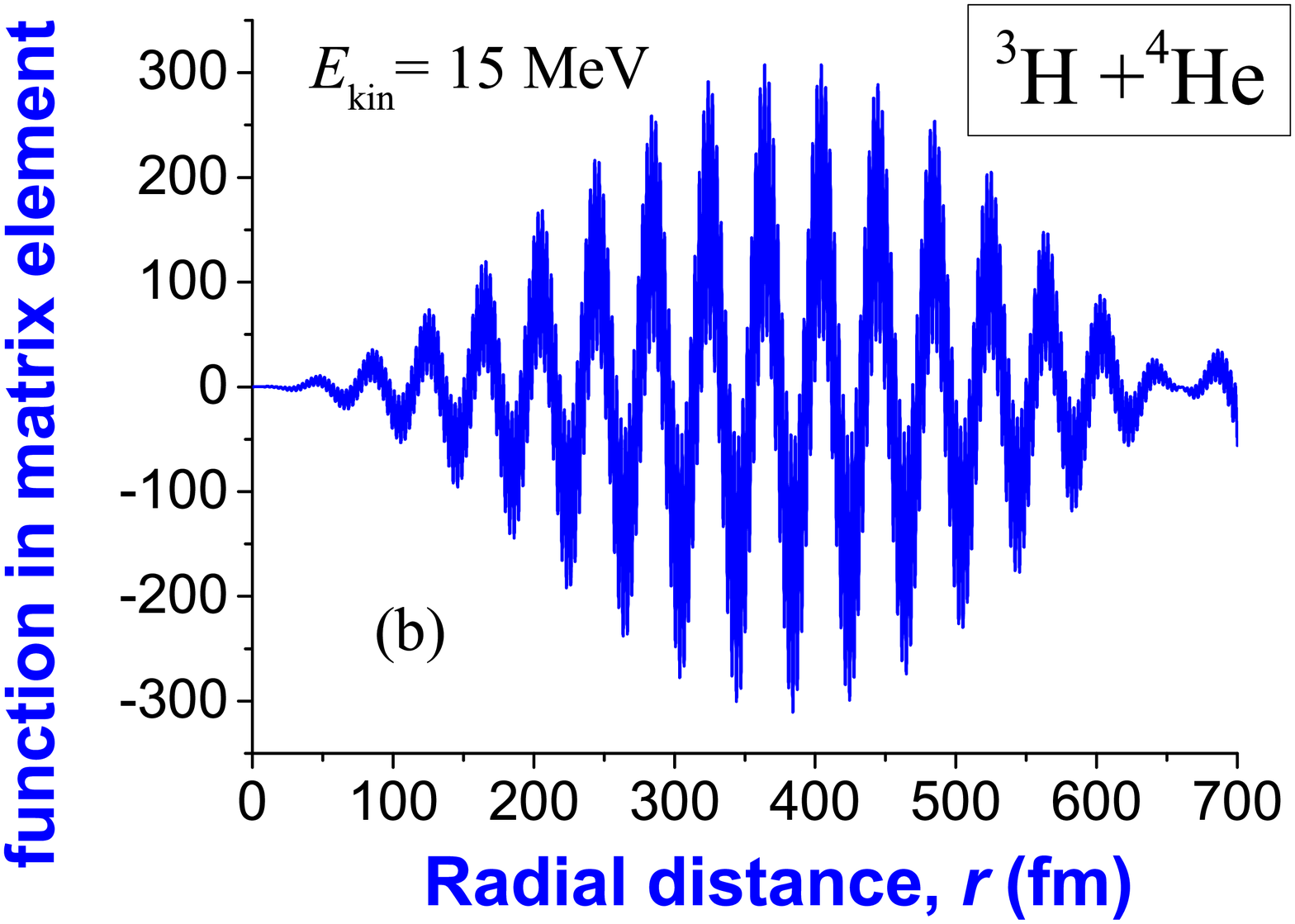}}
\caption{\small (Color online)
Shape of the full integrant function of the bremsstrahlung matrix element for $\isotope[3]{H} + \isotope[4]{He}$ at energy of relative motion of 15~MeV
in the center of mass system
[Parameters of calculations:
energy of photons is chosen as 4~MeV for demonstration
].
\label{fig.1.1}}
\end{figure}

The integrant function of the bremsstrahlung matrix element for $\isotope[3]{H} + \isotope[4]{He}$ at energy of relative motion of 15~MeV is shown in Fig.~\ref{fig.1.1}.
From figures one can clearly see that the radial coordinate region from zero up to 700~fm is the minimum region that includes complete shape of the all harmonics.
However, for calculations of the spectra with minimal satisfactory accuracy it is better to take into account at least a few of these complete shapes.
This specifies the minimum value for $R_{\rm max}$.
Hence, it is clear that this parameter cannot be small.
For calculations in this paper we chose value
$R_{\rm max}= 20000~{\rm fm}$ and 2500000~intervals for radial region of integration.

\subsection{Spectra in dependence on kinetic energy of relative motion between two nuclei
\label{sec.crosssection.2}}

\begin{figure}[htbp]
\centerline{\includegraphics[width=90mm]{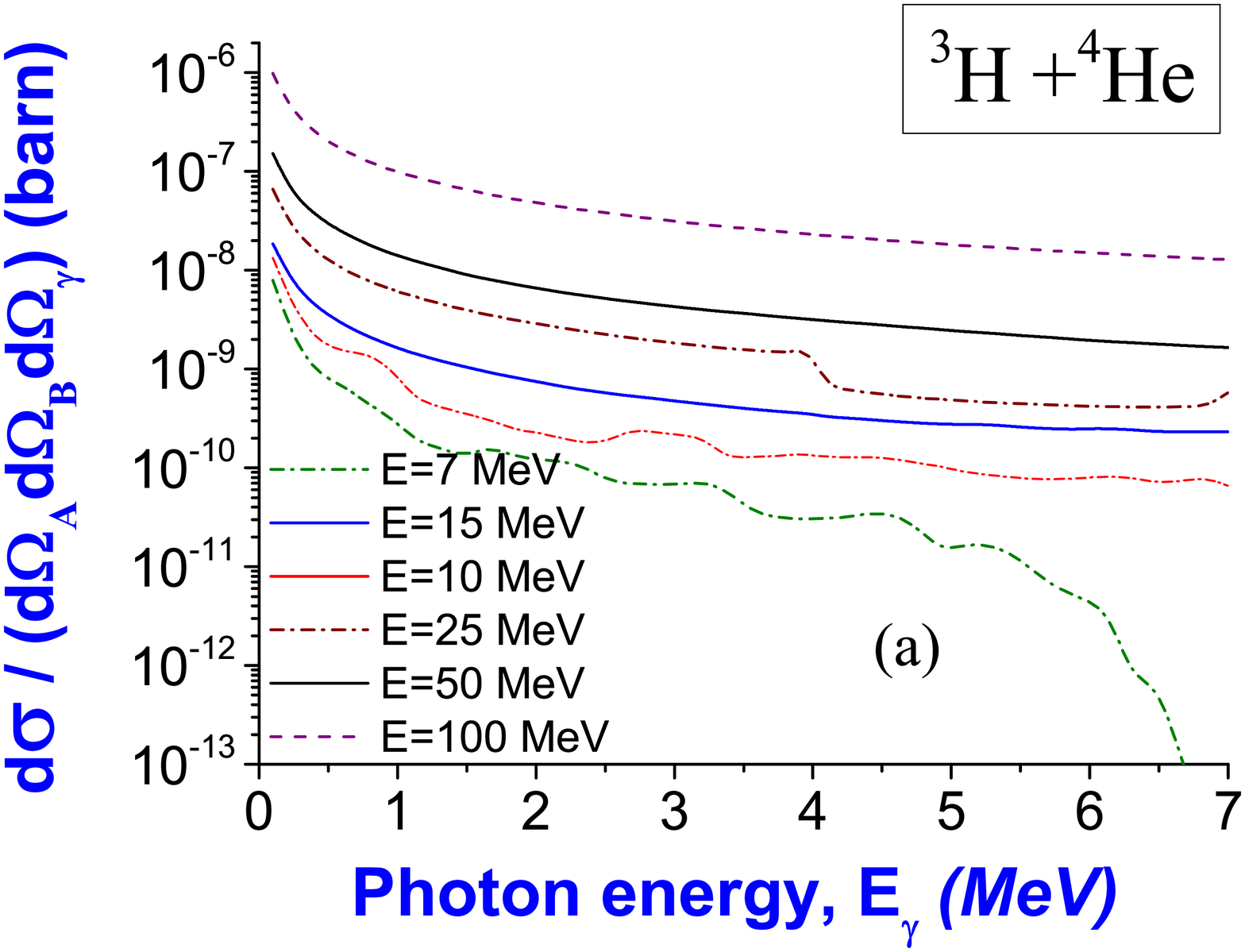}
\hspace{-7mm}\includegraphics[width=90mm]{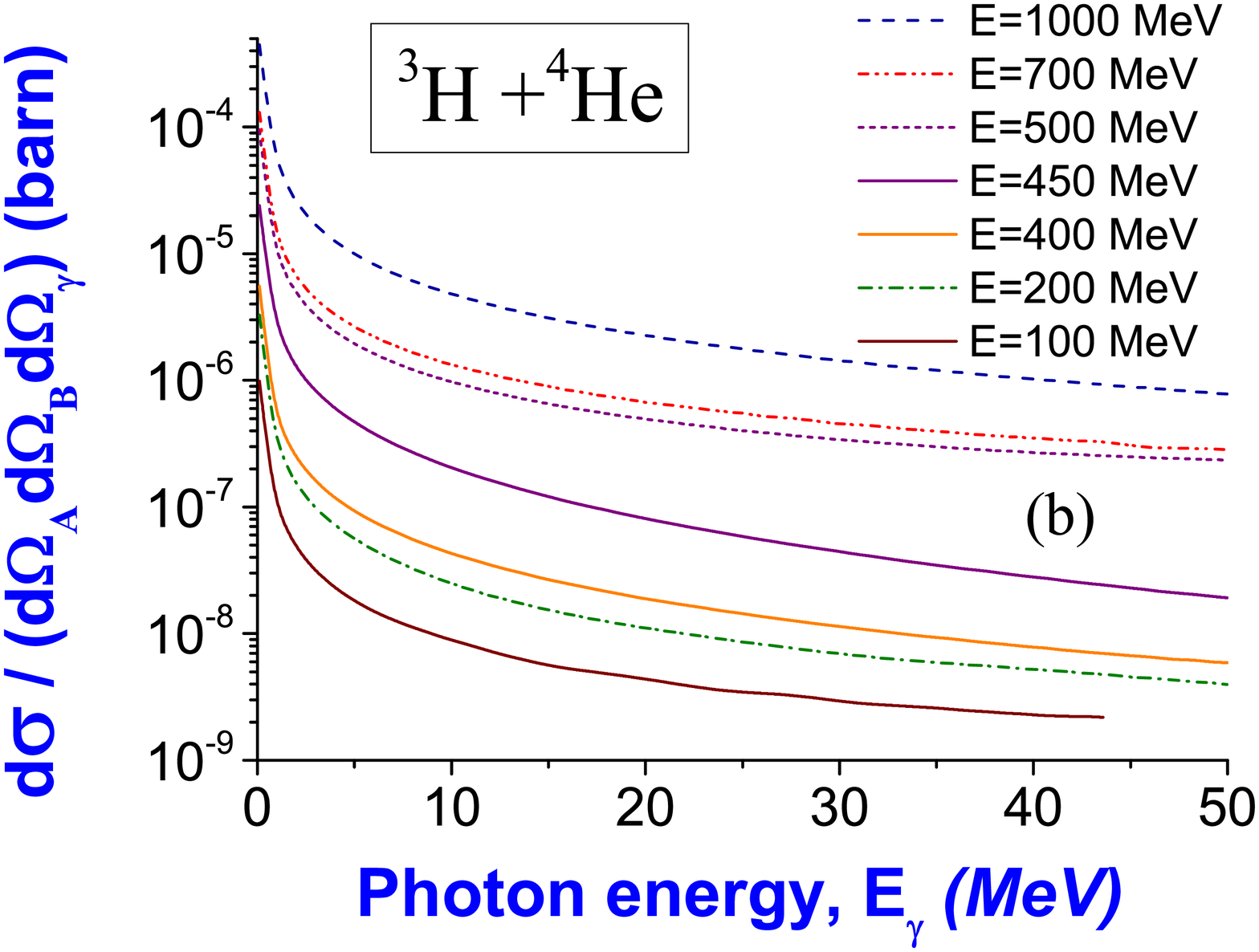}}
\caption{\small (Color online)
Cross sections of bremsstrahlung emission for $^{3}{H} + ^{4}{He}$ at different kinetic energies $E_{kin}$ of relative motion.
[Parameters of calculations:
$R_{max} = 20000$ fm and 2500000~intervals are chosen in the numerical integration.
Time of computer calculation is 53 min for each cross section in figure (a) (50 points for each cross section),
1 hour 12 min for each cross section in figure (b) (70 points for each cross section).]
One can see tendency in each figure that the spectra are increased monotonously at increasing of energy of relative motion $E_{kin}$.
There is essential decreasing of the cross section at increasing of energy of photons for the smallest energy $E_{kin}$ = 7~MeV in figure~(a),
that is explained by tending to the kinematic limit at higher energies of photons.
Rate of increasing of the spectra from energy  $E_{kin}$ is not monotonous, where one can find maximums at some energies.
\label{fig.1.2}}
\end{figure}

At first, let us amalyze how the spectra are chenged in dependence on kinetic energy of relative motion between two nuclei.
In Fig.~\ref{fig.1.2} we present the bremsstrahlung cross sections for $^{3}{H} + ^{4}{He}$ in such a dependence
inside energy range from 7~MeV to 1000~MeV calculated by our approach.

From these figures one can see that our approach gives unified picture of emission of bremsstrahlung photons inside this wide energy region.

\subsection{Cross sections for different nuclei at the same energy of relative motions between nuclei
\label{sec.crosssection.2}}

\begin{figure}[htbp]
\centerline{\includegraphics[width=90mm]{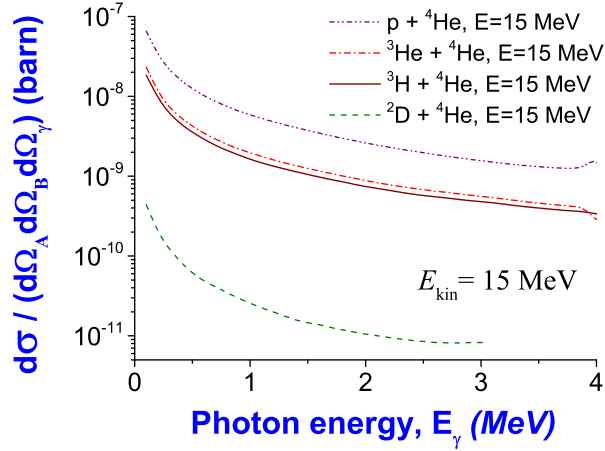}}
\vspace{-3mm}
\caption{\small (Color online)
Cross sections of bremsstrahlung emission for
$p + \isotope[4]{He}$,
$\isotope[2]{D} + \isotope[4]{He}$,
$\isotope[3]{H} + \isotope[4]{He}$,
$\isotope[3]{He} + \isotope[4]{He}$ at kinetic energy of relative motion of $E_{\rm kin}=15$~MeV in center-of-mass frame.
[Parameters of calculations:
cross sections are defined in Eq.~(\ref{eq.crosssection.1.1}),
$R_{\rm max}= 20000\, {\rm fm}$ and 2500000~intervals are chosen in the numerical integration,
time of computer calculation for each cross section is 53~min].
\label{fig.1.3}}
\end{figure}

Cross sections calculated for
$p + \isotope[4]{He}$,
$\isotope[2]{D} + \isotope[4]{He}$,
$\isotope[3]{H} + \isotope[4]{He}$,
$\isotope[3]{He} + \isotope[4]{He}$ at energy 15~MeV are presented in Fig.~\ref{fig.1.3}.
From this figure one can see that (at the same energy $E_{\rm kin}=15$~MeV)
(1) the most intensive emission of photons is for $p + \isotope[4]{He}$,
(2) emissions of photons for $\isotope[3]{H} + \isotope[4]{He}$ and $\isotope[3]{H} + \isotope[4]{He}$ are very similar,
(3) $\isotope[2]{D} + \isotope[4]{He}$ emits photons with the smallest intensity.
One of explanations is in different ratios between form-factors (\ref{eq.folding.2.2}) for these systems
(more precisely,
one can calculate factor
$f =
\sqrt{A_{2} / (A_{1})} \cdot F_{1} -
\sqrt{A_{1} / A_{2}} \cdot F_{2}$ for such estimations).

\subsection{Role of oscillator length $b$ of nuclei in calculations of the spectra
\label{sec.analysis.osc_length}}

\begin{figure}[htbp]
\centerline{\includegraphics[width=90mm]{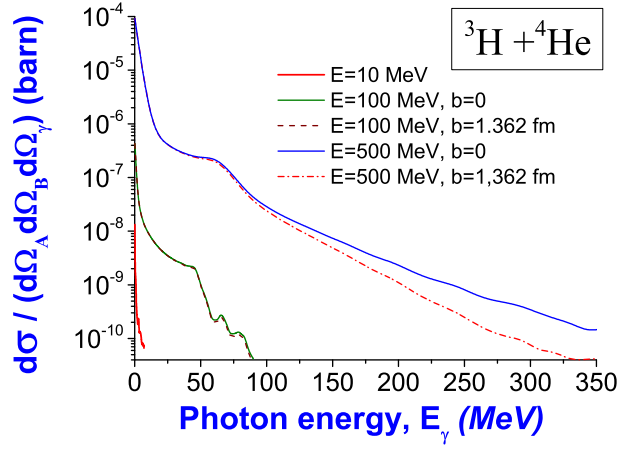}}
\vspace{-3mm}
\caption{\small (Color online)
Cross section of bremsstrahlung emission for $\isotope[3]{H} + \isotope[4]{He}$ at different oscillator lengths $b$
[Parameters of calculations:
cross section is defined in Eq.~(\ref{eq.crosssection.1.1}),
$R_{\rm max}= 10000\, {\rm fm}$ and 5000000~intervals are chosen in the numerical integration].
\label{fig.1.4}}
\end{figure}

Let's clarify how oscillator length influences of the spectra of bremsstrahlung emission.
According to Eqs.~(\ref{eq.simplestcase.1.8}), oscillator length $b$ is included to the matrix element as
\begin{equation}
\begin{array}{llll}
\vspace{2.0mm}
  \Bigl\langle \Psi_{E_{f} l_{f}} \Bigl| \hat{H}_{\gamma} (\vb{k}_{\gamma}, \varepsilonbf_{\mu}) \Bigr| \Psi_{E_{i}l_{i}} \Bigr\rangle =
  -\, i \hbar\,
  \Bigl\{
  \sqrt{\displaystyle\frac{A_{2}}{A_{1}\, A}}\,
    \varepsilonbf_{\mu}\, \vb{I}_{1} \Bigl( +\, \sqrt{\displaystyle\frac{A_{2}}{A_{1}\, A}}\, \Bigr)\, F_{1} -
  \sqrt{\displaystyle\frac{A_{1}}{A_{2}\, A}}\,
    \varepsilonbf_{\mu}\, \vb{I}_{1} \Bigl( -\, \sqrt{\displaystyle\frac{A_{1}}{A_{2}\, A}}\, \Bigr)\, F_{2}
  \Bigr\}, \\

  F_{\alpha}\; =\;
  Z_{\alpha}\, \exp{ - \displaystyle\frac{1}{4}\, \displaystyle\frac{A_{\alpha} - 1}{A_{\alpha}}\, (k, b)^{2}}.
\end{array}
\label{eq.analysis.osc_length.1}
\end{equation}
One can see that in order to study influence of the oscillator length on the spectra,
the exponent in the second formula in Eqs.~(\ref{eq.analysis.osc_length.1}) should be taken into account in calculations.
From this formula one can see that inclusion of the oscillator length suppresses the full bremsstrahlung cross section additionally.
Here a natural question is appeared: how much strong is such a suppression or can it be practically invisible?
A next question is in where is such an influence the most strong?
The simplest way to obtain answer comes from numerical estimations of exponent in dependence on different energies.
More precise information is obtained from calculations of the spectra.
The bremsstrahlung spectra with influence of the oscillator length for different energies are shown in Fig.~~\ref{fig.1.4}.
%
Analyzing such spectra, we conclude the following.
%
\begin{itemize}
\item
Changes of the bremsstrahlung spectra due to change of the oscillator length are practically not visible for energies below 50~MeV.

\item
The oscillator length begins to play visible role for higher energies of the emitted photons (i.e., hundreds of MeV).
This requires to use more large energies of scattering of nuclei.
Such highest sensitivity in cross sections at higher energies can be explained by the following.
Each matrix element of bremsstrahlung emission includes two wave functions of relative motion, i.e. these are wave functions for states before emission of photon and after such an emission.
Wave function before emission of photon is defined concerning to higher energy of relative motion $E_{\rm kin}$,
so its wave length is shorter that helps to distinguish more details (more tiny microstructure) in the shape of potential with barrier.
Wave function after emission of higher energetic photon is defined concerning to smaller energy $E_{\rm kin}$.
That allows to analyze underbarrier tunneling effects and shape of barrier.
In the matrix element properties of these two wave functions are combined.
Indeed, this is visible Fig.~\ref{fig.1.4} that confirms such a logic.

\item
There is no sense to analyze spectra at low energies (energies of photons and energies of relative motion between two incident nuclei).
Instead of this, it needs to choose window in high energy region for such analysis.
In this place cross sections are smaller essentially, that makes to be more difficult experimental measurements and theoretical calculations.
This aspect gives own restriction on maximally large energy, higher that study of this question will be not realistic (impossible).
%
\end{itemize}

\subsection{Is it possible to see in the spectra the influence of the nuclear component of the interaction potential and where?
\label{sec.analysis.osc_length}}

\begin{figure}[htbp]
\centerline{\includegraphics[width=90mm]{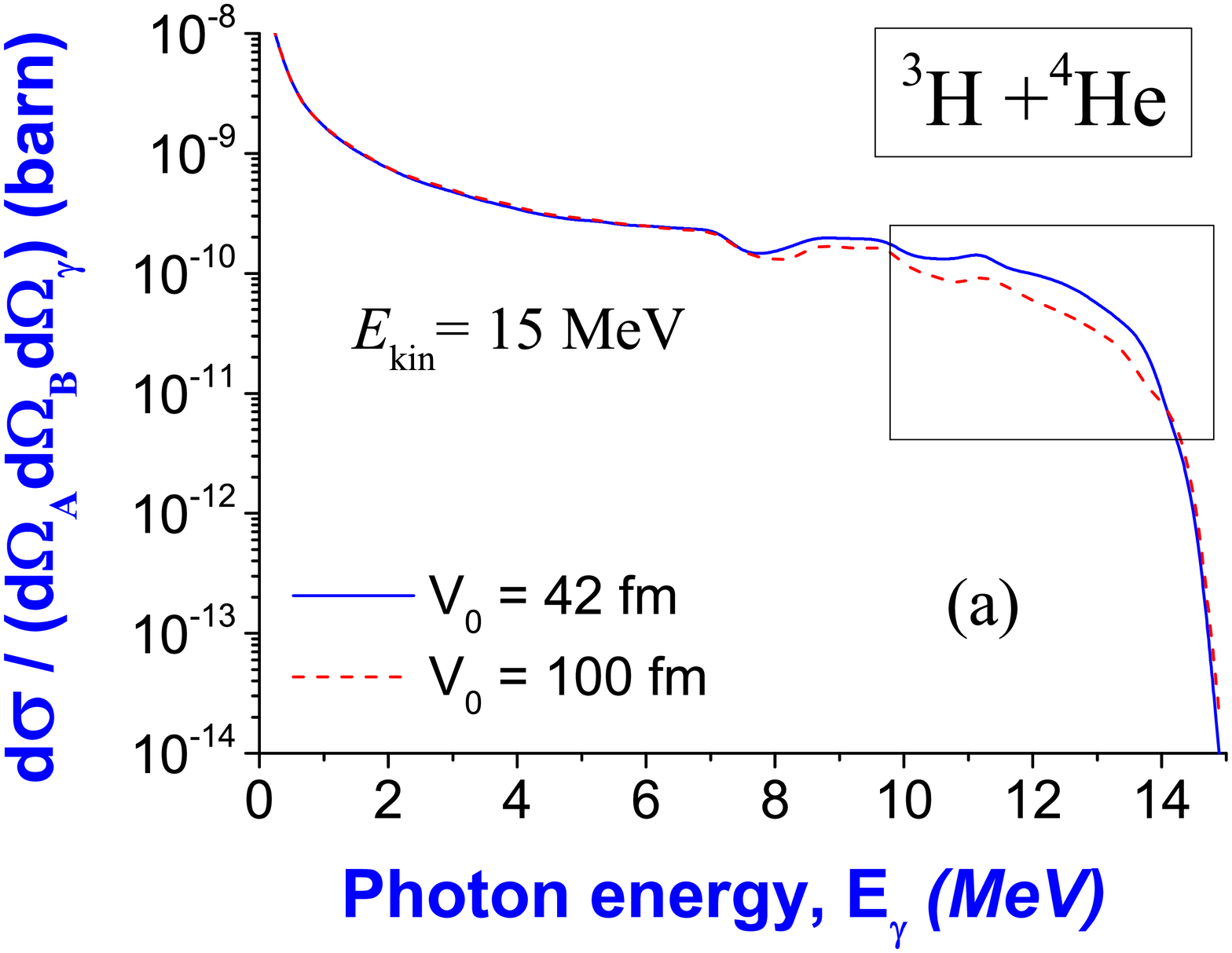}
\hspace{-8mm}\includegraphics[width=90mm]{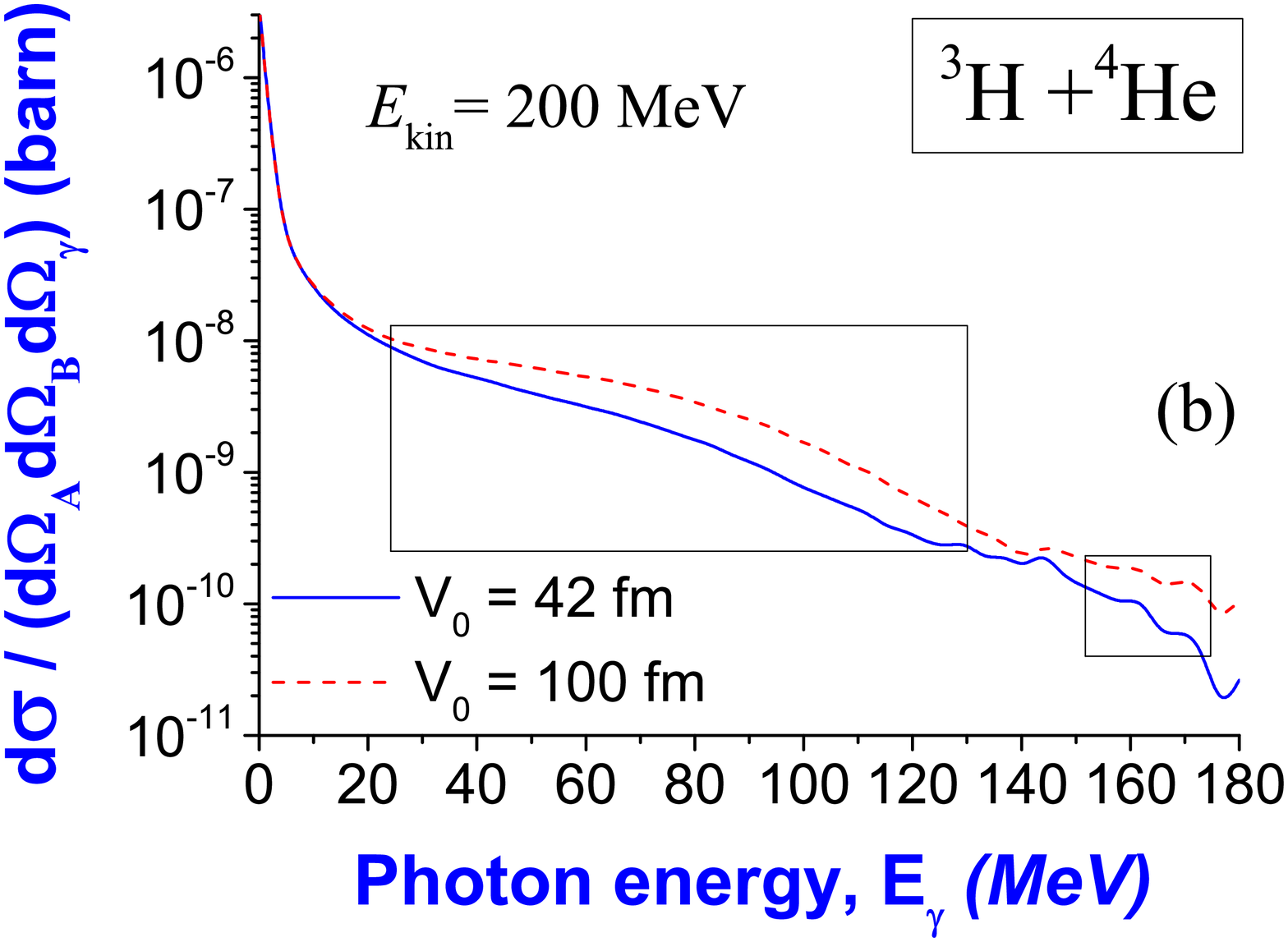}}
\vspace{-2mm}
\caption{\small (Color online)
Cross sections of bremsstrahlung emission for $\isotope[3]{H} + \isotope[4]{He}$ in dependence on depth of nuclear potential $V_{0}$
at energy of relative motion of 15~MeV (a) and 200~MeV~(b) in center-of-mass frame
[Parameters of calculations:
cross section is defined in Eq.~(\ref{eq.crosssection.1.1}),
$R_{\rm max}= 2000\, {\rm fm}$ and 2500000~intervals are chosen in the numerical integration].
\label{fig.1.5}}
\end{figure}

Let us analyze whether it is possible to see in the spectra of photon emission the influence of parameters of the nuclear part of the interaction potential.
Calculations of the spectra for $\isotope[3]{H} + \isotope[4]{He}$ in dependence on the depth of the nuclear component of the potential are shown in Fig.~\ref{fig.1.5}.
%
In the figure (a), calculations are given at $E_{\rm kin} = 15$~MeV, where one can chose a region in the range of higher photon energies (10--14~MeV), where there is small visible difference between the spectra (this area in the figure is highlighted with a rectangle).
One can see change in the spectra in dependence on the depth of the nuclear part $V_{0}$ of the interaction potential.
It would seem that this picture should be good indication of the place where it is worth looking for the influence of the parameters of the nucleus part of the potential on the bremsstrahlung cross section:
\emph{This is a region with higher photon energies (at any energy of relative motion between nuclei)}.
%

However, let us analyze whether such a dependence will persist with an increase in the energy of the relative motion of the nuclei.
The spectra at energy of relative motion of nuclei of 200 MeV are presented in Fig.~\ref{fig.1.5}~(b).
Here there is another region at range of 25--130~MeV, where difference between the spectra is clearly visible.
Note that this type of dependence has never been found yet in study on bremsstrahlung emission in nuclear reactions.
%
%
Such a dependence is likely to have opposite character than the previous dependence revealed in Fig.~\ref{fig.1.5}~(a).
It seems, this dependence is more suitable for experimental investigations.
Because it allows use of essentially lower photon energies for measurements.
This dependence is much more reliable (i.e., it is shown in a wider region of photon energies).
%
%
The spectra have more intense emission, that is easier to measure and gives higher convergence in computer calculations.
If now, after such a detection of dependence at $E_{\rm kin} = 200$~MeV, to go back to Fig.~\ref{fig.1.5}~(a) at $E_{\rm kin} = 15$~MeV,
then one can find a suitable difference between the spectra of a similar nature for lower photon energies (at 2--6~MeV).
This reinforces confidence in the analysis obtained.
%

\subsection{Analysis of experimental bremsstrahlung data
\label{sec.expanalysis}}



%







\subsubsection{Bremsstrahlung in the proton-deuteron scattering
\label{sec.expanalysis.1}}

\begin{figure}[htbp]
\centerline{\includegraphics[width=90mm]{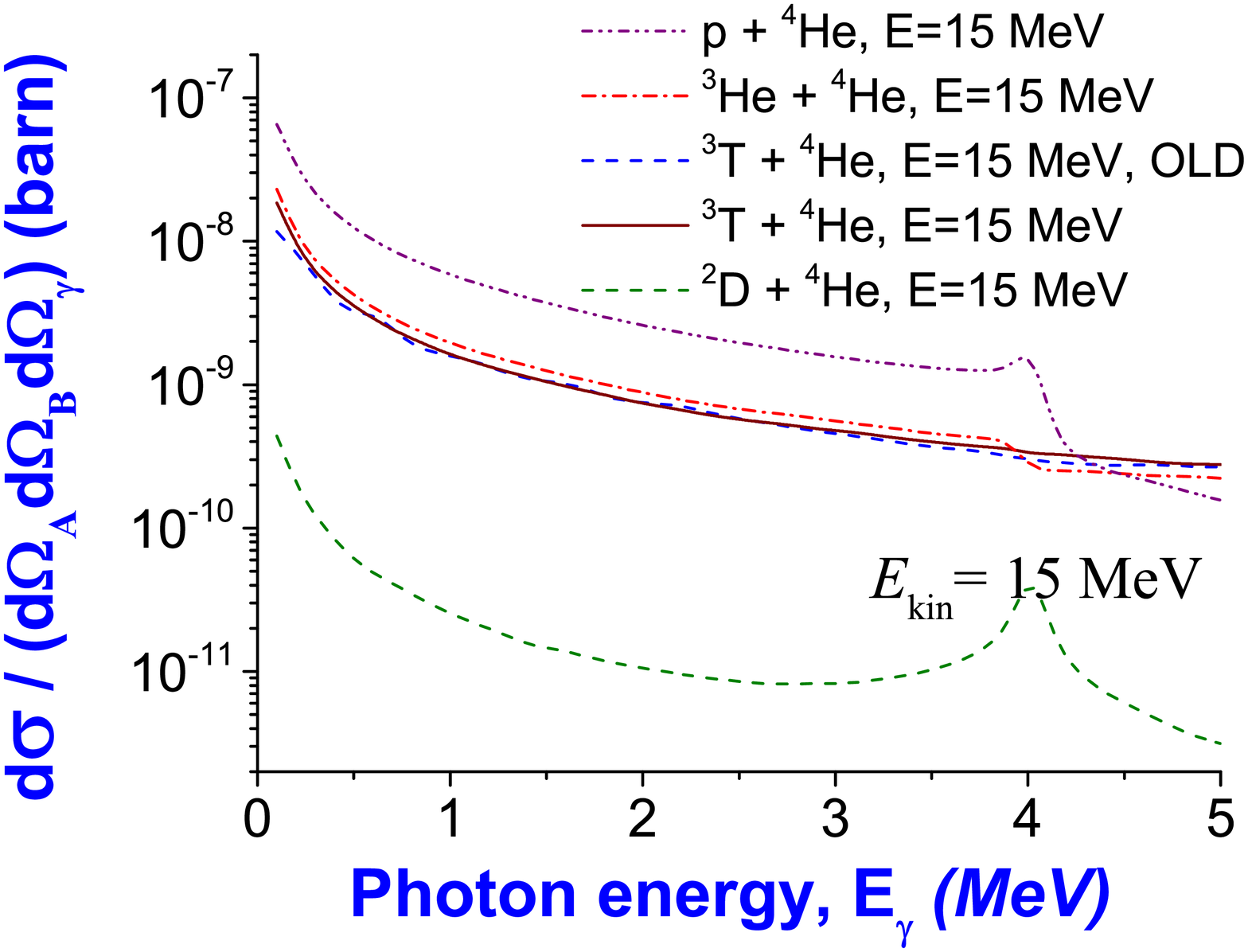}}
\vspace{-2mm}
\caption{\small (Color online)
Normalized cross sections of bremsstrahlung emission for $p + \isotope[2]{D}$
at energy of proton beam of 145~MeV
in comparison with experimental data \cite{Clayton.1992.PRC.p1810}
[Parameters of calculations:
cross section is defined in Eq.~(\ref{eq.crosssection.1.1}),
$R_{\rm max}= 20000\, {\rm fm}$ and 2500000~intervals are used in the numerical integration;
kinetic energy $E_{\rm kin}$ of relative motion of proton and deuteron is used in calculations of bremsstrahlung matrix elements,
which is $E_{\rm kin} = 2/3 \cdot E_{\rm p} = 98$~MeV].
Bremsstrahlung calculation of Herrmann, Speth and Nakayama for $p + n$ at 150~MeV~\cite{Herrmann.1991.PRC} is added for comparative analysis
(see the red dashed line in figure).
\label{fig.exp.1}}
\end{figure}

We will analyze bremsstrahlung from the proton-deuteron scattering.
There are different aspects of use of angles in calculations of the bremsstrahlung cross sections.
Moreover, in the different papers authors sometimes used re-normalized experimental data for that reaction.
But, one can obtain understanding about general tendency of bremsstrahlung cross section from normalized calculations.
By such a reason, we will provide normalized calculations of spectra in this paper.
Calculations of bremsstrahlung spectra on the basis of our model in comparison with experimental data of Clayton et al. \cite{Clayton.1992.PRC.p1810}
are presented in Fig.~\ref{fig.exp.1}.
From these figures one can see that our approach describes these data in Ref.~\cite{Clayton.1992.PRC.p1810} with enough good agreement.
%
One can suppose that one can improve agreement between our calculations and experimental data~\cite{Clayton.1992.PRC.p1810} in the middle part of the spectra, if to add influence of magnetic moments of nucleons (related with spins of these nucleons) on emission of photons.

For comparative analysis we add also calculation~\cite{Herrmann.1991.PRC} of Herrmann, Speth and Nakayama, which was used in analysis in Ref.~~\cite{Clayton.1992.PRC.p1810}.
After renormalization, general tendency of the spectra are not changed actually (differences are very small).
By such a reason, we think that our calculation is in better agreement with experimental data~\cite{Clayton.1992.PRC.p1810}
than calculation~\cite{Herrmann.1991.PRC} (see red dashed line in figure).

\subsubsection{Bremsstrahlung in the scattering of the $\alpha$-particles on protons
\label{sec.expanalysis.2}}

In this section we will analyze bremsstrahlung emission in the scattering of $\alpha$~particles on protons.
Photons of bremsstrahlung were measured over large range for the reaction of $\alpha$ particles with protons,
using photon spectrometer TAPS at AGOR facility of the Kernfysisch Verneller Institut \cite{Hoefman.2000.PRL}.
In this experiment the beam of 200~MeV $\alpha$ particles was incident on a liquid hydrogen target.
Experimental bremsstrahlung data for this study were presented in papers~\cite{Hoefman.2000.PRL,Hoefman.1999.NPA}
and PhD thesis \cite{Hoefman.1999.PhD}.

From these presented data we choose data in Fig.~3 in Ref.~\cite{Hoefman.2000.PRL} for analysis.
Note that inclusive data given in Fig.~1 and exclusive data given in Fig.~3 in that paper are different at $E_{\gamma} < 20$~MeV
(with smaller difference at $E_{\gamma} > 20$~MeV).
We explain our choice by the following.
(1) Data in Fig.~3 in Ref.~\cite{Hoefman.2000.PRL} were obtained on the basis of double and triple coincidences.
(2) On the basis of these data authors of that paper obtained information about resonances (radiative capture populating the unbound ground and first excited states) of shortly lived nucleus \isotope[5]{Li}, authors extracted parameters presented in Tabl.~1 in Ref.~\cite{Hoefman.2000.PRL}, and calculated the bremsstrahlung contributions for these resonances shown in Figs.~1 in that paper.

\begin{figure}[htbp]
\centerline{\includegraphics[width=90mm]{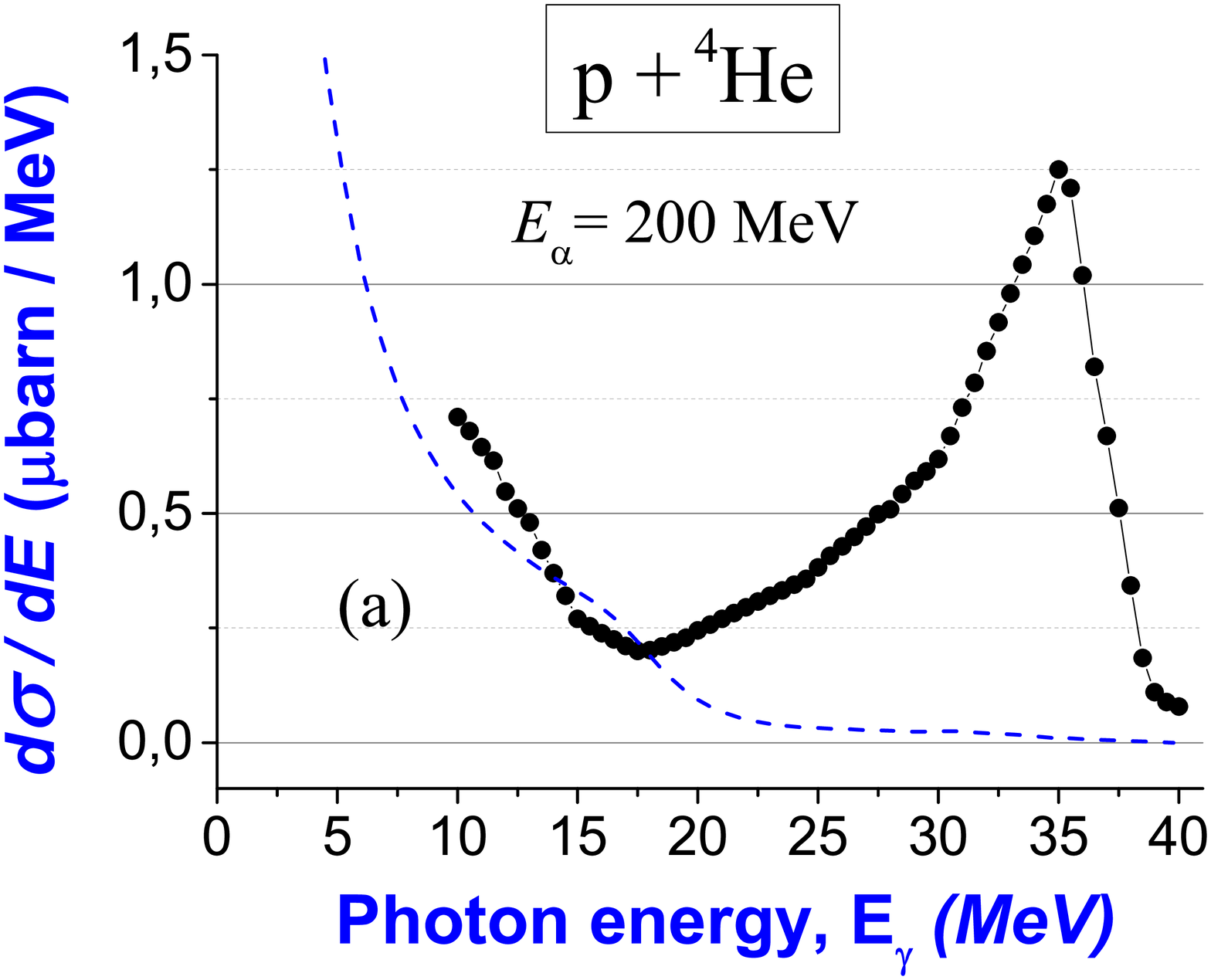}
\hspace{-7mm}\includegraphics[width=90mm]{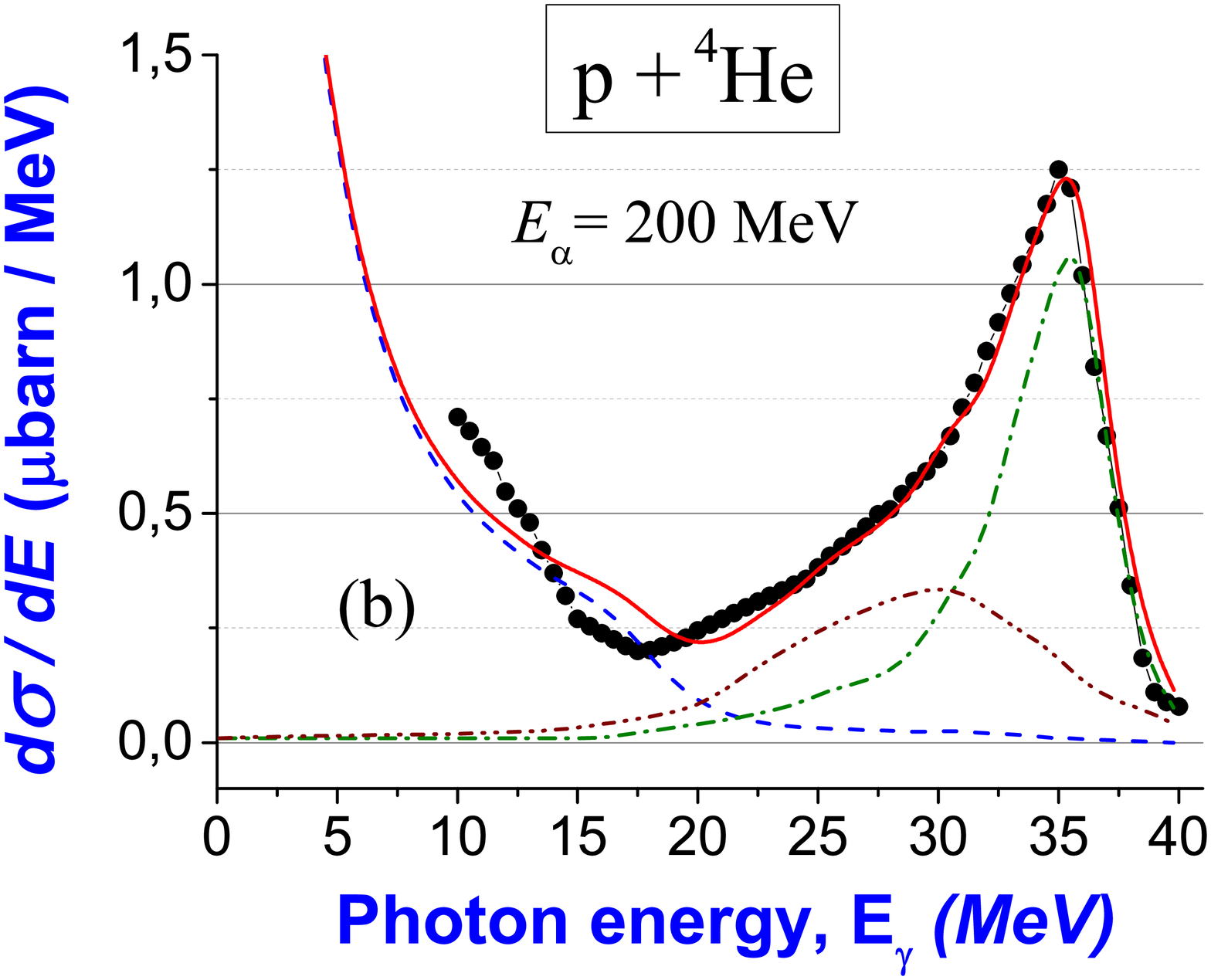}}
\vspace{-2mm}
\caption{\small (Color online)
Calculated cross sections of bremsstrahlung emission for $p + \isotope[4]{He}$
at energy of beam of the $\alpha$~particles of 200~MeV
in comparison with experimental data~\cite{Hoefman.2000.PRL}
[Parameters of calculations:
cross section is defined in Eq.~(\ref{eq.crosssection.1.2}),
$R_{\rm max}= 20000\, {\rm fm}$ and 2500000~intervals are used in the numerical integration%
].
[Pannel (a)]:
Main bremsstrahlung contribution of photons emitted in the $p + \alpha$ scattering calculated by our model (see blue dashed line)
in comparison with experimental data.
[Pannel (b)]:
Full bremsstrahlung cross section (see red solid line) obtained as summation of
main contribution in the $p + \alpha$ scattering (see blue dashed line)
and additional two contributions from capture at
unbound ground and first excited states of \isotope[5]{Li}
(see green dash-dotted line and brown dash-double dotted line)
in comparison with experimental data.
\label{fig.exp.2}}
\end{figure}

As it was shown in Ref.~\cite{Hoefman.2000.PRL},
for explanation of experimental data it needs to take into account
presence of two unbound states of \isotope[5]{Li} for capture
in addition to the main bremsstrahlung emission during the $p + \alpha$ scattering
(it is clearly seen in Figs.~1 and 3 in that paper).
In particular, in Fig.~1 in Ref.~\cite{Hoefman.2000.PRL} such states were obtained as two Gaussian peaks.
Parameters of these Gaussian peaks were deduced using fitting procedure (see Tabl.~1 in that paper).
In our paper we do not take into account incoherent part and magnetic part of bremsstrahlung.
So, our approach reproduces main contribution during scattering,
but to take into account two resonant states of \isotope[5]{Li}
we follow to results of research~\cite{Hoefman.2000.PRL}.
Calculations of bremsstrahlung spectra on the basis of our model in comparison with experimental data of Hoefman et al.~\cite{Hoefman.2000.PRL}
are presented in Fig.~\ref{fig.exp.2}.

Our model calculates main bremsstrahlung contribution on the microscopic basis,
while in Ref.~\cite{Hoefman.2000.PRL} this contribution was obtained on the basis of phenomenological formula~(1) from classical electrodynamics with energy and momentum conservations.
The summarized full cross section from our calculations is shown in Fig.~\ref{fig.exp.2}~(b) by red solid line.
One can see that agreement between such our calculated full cross section and experimental data~\cite{Hoefman.2000.PRL} is comparable with result in Ref.~\cite{Hoefman.2000.PRL}.
Also we add the full bremsstrahlung cross section at low energy region of photons (in contrast to Ref.~\cite{Hoefman.2000.PRL}).%
\footnote{%
Presence of smoothly visible hump in the calculated cross section at photon energy close to 16~MeV can be explained by restriction of numerical calculation of Coulomb functions in the close asymptotic region.
Note that Coulomb functions at some parameters are calculated on the basis of asymptotic series.
These series are not convergent, in principle.
But the best calculated values have not bad approximation and they are generally accepted in physical community for use.
However, this forms origin of small changes in the full bremsstrahlung spectra at some parameters.
We suppose accuracy of calculation of this part of the cross section can be improved on the basis of technique developed in Appendix in Ref.~\cite{Maydanyuk.2010.PRC}.
However, this technically solid volume of development is omitted in this paper,
and we restrict ourselves by current existed accuracy of these functions.}

Note that for analysis authors of research~\cite{Hoefman.2000.PRL} also used
the potential model of Baye \emph{et al.}~\cite{1992NuPhA.550..250B},
the covariant generalization of the Feshbach-Yennie approach based on Refs.~\cite{Ding.1989.PRC,Liou.1995.PLB}, and
model of direct capture into unbound states based on Ref.~\cite{Weller.1982.PRC}
(see Fig.~3 in Ref.~\cite{Hoefman.2000.PRL}).
However, at low energy region of the bremsstrahlung cross section these approaches are less successful than our microscopic cluster model.

\section{Conclusions and perspectives
\label{sec.conclusions}}

In this paper a new model of bremsstrahlung emission in the scattering of light nuclei is constructed with main focus on strict cluster formulation of nuclear processes.
But, analysis is performed in frameworks of the folding approximation of the formalism with participation of $s$-nuclei.
%
%
On the basis of such a model we investigate emission of the bremsstrahlung photons in the scattering of the light nuclei.
Conclusions from such a study are the following.
\begin{itemize}

\item
On the example of $\isotope[3]{H} + \isotope[4]{He}$, we obtain unified picture of emission of bremsstrahlung photonsn in dependence on
the different kinetic energies $E_{\rm kin}$ of relative motion of two nuclei inside wide energy range from 7~MeV to 1000~MeV calculated by our approach (see Fig.~\ref{fig.1.2}).
Cross sections are increased at increasing of energy of relative motion $E_{\rm kin}$.
But, rate of increasing of the spectra from energy  $E_{\rm kin}$ is not monotonous, where one can find maximums at some energies.

\item
We estimate comparable picture of the bremsstrahlung emission for
$p + \isotope[4]{He}$,
$\isotope[2]{D} + \isotope[4]{He}$,
$\isotope[3]{H} + \isotope[4]{He}$,
$\isotope[3]{He} + \isotope[4]{He}$
at the same kinetic energy of relative motion of $E_{\rm kin}=15$~MeV (see Fig.~\ref{fig.1.3}).
General tendency of the spectra is similar for all systems.
We find that
(1) the most intensive emission of photons is for $p + \isotope[4]{He}$,
(2) emissions of photons for $\isotope[3]{H} + \isotope[4]{He}$ and $\isotope[3]{He} + \isotope[4]{He}$ are very similar,
(3) $\isotope[2]{D} + \isotope[4]{He}$ emits photons with the smallest intensity.

\item
We have analyzed influence of the oscillator length $b$ on the spectra of bremsstrahlung emission
[see Eqs.~(\ref{eq.analysis.osc_length.1}), Fig.~\ref{fig.1.4}] and conclude the following.
%
%
%
\begin{itemize}
\item
Inclusion of the oscillator length suppresses the full bremsstrahlung cross section additionally.

\item
Changes of the bremsstrahlung spectra due to change of the oscillator length are practically not visible for energies below 50~MeV.

\item
The oscillator length begins to play visible role for higher energies of the emitted photons (i.e., hundreds of MeV).
This requires to use more large energies of the incident nuclei.

\item
There is no sense to analyze spectra at low energies.
Instead of this, it needs to choose window in high energy region for such analysis.
In this place cross sections are smaller essentially, that makes to be more difficult experimental measurements and theoretical calculations.
This aspect gives restriction on maximally large energy, above that study of this question will be not realistic (impossible).
\end{itemize}

\item
We have found dependence of the bremsstrahlung spectra on parameters of the nuclear part of the interaction potential.
On the example of $\isotope[3]{H} + \isotope[4]{He}$,
we find two regions in the spectra for better observation of such a dependence: the middle energy region and high energy region (see Fig.~\ref{fig.1.5}).
In the middle energy region calculations of the spectra are more stable, intensity of emission is higher that is convenient for possible measurements.
In the high energy region dependence is more sensitive to variation of nuclear parameters. But intensity of emission is smaller that is problematic for possible measurements and computer calculations.

\item
The bremsstrahlung cross section calculated on the basis of our model for the proton-deuteron scattering at $E_{p}=145$~MeV is in good agreement with experimental data Ref.~\cite{Clayton.1992.PRC.p1810} (see Fig.~\ref{fig.exp.1}).

\item
The bremsstrahlung emission for the scattering of $\alpha$ particles on protons at energy of beam of $\alpha$ particles of $E_{\alpha}=200$~MeV is analyzed.
The bremsstrahlung cross section obtained as summation of
main contribution in the $p + \alpha$ scattering (calculated on the basis of our microscopic model)
and additional two contributions from captures at
unbound ground and first excited states of \isotope[5]{Li} is in good agreement with experimental data~\cite{Hoefman.2000.PRL} (see Fig.~\ref{fig.exp.2}).

\end{itemize}
As a perspective, we estimate useful new contribution from inclusion of magnetic moments of nucleons to the model.
This should give additional new incoherent bremsstrahlung contribution.
According to previous study of bremsstrahlung in the scattering of protons of heavy nuclei \cite{Maydanyuk_Zhang.2015.PRC}, this incoherent emission is important and not small.
However, its role can be essentially smaller for light nuclei, according to our preliminary estimations.
We also estimate that this term can improve agreement between the full calculated cross section and experimental data.


\section*{Acknowledgements
\label{sec.acknowledgements}}

M.S.P. thanks the Wigner Research Centre for Physics in Budapest for warm hospitality. 
This work was supported in part by the Program of Fundamental Research of the
Physics and Astronomy Department of the National Academy of Sciences of
Ukraine (Project No. 0117U000239).

\appendix
\section{Matrix element in the folding approximation
\label{sec.app.folding}}

Let us consider the operator
%
\begin{equation}
  \widehat{H}_{e}\left( \mathbf{k}_{\gamma },\varepsilonbf_{\mu}\right) =
  \frac{e\hbar }{m_{N}c}\sum_{i=1}^{A}\frac{1}{2}\left( 1+\widehat{\tau }_{iz}\right)
  \mathbf{A}^{\ast }\left( i\right) \widehat{\pibf}_{i}^{\ast}
\label{eq.app.folding.1.1}
\end{equation}
for two-cluster system with the partition $A = A_{1} + A_{2}$. In this case the operator
$\widehat{H}_{e}\left( \mathbf{k}_{\gamma },\varepsilonbf_{\mu }\right) $ can be presented as
\begin{equation}
\begin{array}{lll}
\vspace{1.5mm}
  \widehat{H}_{e}\left( \mathbf{k}_{\gamma },\varepsilonbf_{\mu}\right)  & = &
  \widehat{H}_{e}^{\left( 1\right) }\left( \mathbf{k}_{\gamma },\varepsilonbf_{\mu }\right) +
  \widehat{H}_{e}^{\left( 2\right)}\left( \mathbf{k}_{\gamma },\varepsilonbf_{\mu }\right) \\

  & = &
  \displaystyle\frac{e\hbar }{m_{N}c}\sum_{i\in A_{1}}
  \displaystyle\frac{1}{2}\left( 1+\widehat{\tau }_{iz}\right)
  \mathbf{A}^{\ast }\left( i\right) \widehat{\pibf}_{i}^{\ast } +
  \displaystyle\frac{e\hbar }{m_{N}c}\sum_{j\in A_{2}}\frac{1}{2}\left( 1 + \widehat{\tau }_{jz}\right)
  \mathbf{A}^{\ast }\left( j\right) \widehat{\pibf}_{j}^{\ast }.
\end{array}
\label{eq.app.folding.1.2}
\end{equation}
%
It is necessary to recall that
\begin{equation}
\begin{array}{lll}
  \mathbf{A}^{\ast }\left( i\right) \widehat{\pibf}_{i}^{\ast } & = &

  \exp \left\{ -i\left( \mathbf{k}_{\gamma }\rhobf_{i}\right)
  \right\} \left( \varepsilonbf_{\mu }\widehat{\pibf}_{i}^{\ast }\right) \\

  & = &
  \exp \left\{ -i\left( \mathbf{k}_{\gamma },\mathbf{r}_{i}-\mathbf{R}_{cm}\right) \right\}
  \left( \varepsilonbf_{\mu }
  \left[ \widehat{\mathbf{p}}_{i}^{\ast }-\widehat{\mathbf{P}}_{cm}^{\ast }\right] \right).
\end{array}
\label{eq.app.folding.1.3}
\end{equation}
The first operator
$\widehat{H}_{e}^{\left( 1\right) }\left( \mathbf{k}_{\gamma },\varepsilonbf_{\mu }\right)$
we represent in the following form
\begin{equation}
\begin{array}{lll}
  \widehat{H}_{e}^{\left( 1\right) }\left( \mathbf{k}_{\gamma },\varepsilonbf_{\mu }\right) & = &




  \displaystyle\frac{e\hbar }{m_{N}c}\exp \left\{ -i\left( \mathbf{k}_{\gamma },\mathbf{R}_{1}-\mathbf{R}_{cm}\right) \right\}

  \displaystyle\sum_{i\in A_{1}}\frac{1}{2}\left( 1+\widehat{\tau }_{iz}\right)
  \exp \left\{ -i\left( \mathbf{k}_{\gamma },\mathbf{r}_{i}-\mathbf{R}_{1}\right) \right\}
  \left( \varepsilonbf_{\mu }\left[ \widehat{\mathbf{p}}_{i}^{\ast }-\widehat{\mathbf{P}}_{1}^{\ast }\right] \right) \\

  & + &
  \displaystyle\frac{e\hbar }{m_{N}c}\exp \left\{ -i\left( \mathbf{k}_{\gamma },\mathbf{R}_{1}-\mathbf{R}_{cm}\right) \right\}
  \left( \varepsilonbf_{\mu }\left[ \widehat{\mathbf{P}}_{1}^{\ast }-\widehat{\mathbf{P}}_{cm}^{\ast }\right] \right)

  \displaystyle\sum_{i\in A_{1}}\frac{1}{2}\left( 1+\widehat{\tau }_{iz}\right)
  \exp \left\{ -i\left( \mathbf{k}_{\gamma },\mathbf{r}_{i}-\mathbf{R}_{1}\right) \right\},
\end{array}
\label{eq.app.folding.1.4}
\end{equation}
where $\mathbf{R}_{1}$ and $\widehat{\mathbf{P}}_{1}^{\ast }$ are coordinate
and momentum of center of mass motion of the first cluster:
\begin{equation}
\begin{array}{lll}
  \mathbf{R}_{1} = \displaystyle\frac{1}{A_{1}} \displaystyle\sum_{i\in A_{1}}\mathbf{r}_{i}, &
  \widehat{\mathbf{P}}_{1} = \displaystyle\frac{1}{A_{1}} \displaystyle\sum_{i\in A_{1}}\widehat{\mathbf{p}}_{i}.
\end{array}
\label{eq.app.folding.1.5}
\end{equation}
%
In similar way we can present the second operator
\begin{equation}
\begin{array}{lll}
  \widehat{H}_{e}^{\left( 2\right) }\left( \mathbf{k}_{\gamma },\varepsilonbf_{\mu }\right) = 

  & = &
  \displaystyle\frac{e\hbar}{m_{N}c}\exp \left\{ -i\left( \mathbf{k}_{\gamma },\mathbf{R}_{2}-\mathbf{R}_{cm}\right) \right\} 

  \displaystyle\sum_{j\in A_{2}}\frac{1}{2}\left( 1+\widehat{\tau }_{jz}\right)
  \exp \left\{ -i\left( \mathbf{k}_{\gamma },\mathbf{r}_{j}-\mathbf{R}_{2}\right) \right\}
  \left( \varepsilonbf_{\mu }\left[ \widehat{\mathbf{p}}_{j}^{\ast }-\widehat{\mathbf{P}}_{2}^{\ast }\right] \right) \\

  & + &
  \displaystyle\frac{e\hbar}{m_{N}c}\exp \left\{ -i\left( \mathbf{k}_{\gamma },\mathbf{R}_{2}-\mathbf{R}_{cm}\right) \right\}
  \left( \varepsilonbf_{\mu }\left[ \widehat{\mathbf{P}}_{2}^{\ast }-\widehat{\mathbf{P}}_{cm}\right]
  \right) 

  \displaystyle\sum_{j\in A_{2}}
  \displaystyle\frac{1}{2}\left( 1+\widehat{\tau }_{jz}\right)
  \exp \left\{ -i\left( \mathbf{k}_{\gamma },\mathbf{r}_{j}-\mathbf{R}_{2}\right) \right\} ,
\end{array}
\label{eq.app.folding.1.6}
\end{equation}
%
where
\begin{equation}
\begin{array}{lll}
  \mathbf{R}_{2} = \displaystyle\frac{1}{A_{2}} \displaystyle\sum_{j\in A_{2}}\mathbf{r}_{j}, &
  \widehat{\mathbf{P}}_{2} = \displaystyle\frac{1}{A_{2}} \displaystyle\sum_{j\in A_{2}}\widehat{\mathbf{p}}_{j}.
\end{array}
\label{eq.app.folding.1.7}
\end{equation}
%
Consider in detail the differences
\begin{equation}
\begin{array}{lll}
  \mathbf{R}_{1}-\mathbf{R}_{cm}, &
  \mathbf{R}_{2}-\mathbf{R}_{cm}.
\end{array}
\label{eq.app.folding.1.8}
\end{equation}
%
Taking into account their definitions, we obtain
\begin{equation}
\begin{array}{lll}
  \mathbf{R}_{1}-\mathbf{R}_{cm} =
  \displaystyle\frac{1}{A_{1}} \displaystyle\sum_{i\in A_{1}} \mathbf{r}_{i} -
  \displaystyle\frac{1}{A} \displaystyle\sum_{i=1}^{A}\mathbf{r}_{i} = 




  \sqrt{\displaystyle\frac{A_{2}}{A_{1}A}} \sqrt{\displaystyle\frac{A_{1}A_{2}}{A}}\left[ \frac{1}{A_{1}}
  \displaystyle\sum_{i\in A_{1}}\mathbf{r}_{i} -
  \displaystyle\frac{1}{A_{2}} \displaystyle\sum_{j\in A_{2}} \mathbf{r}_{j}\right], \\

  \mathbf{R}_{2} - \mathbf{R}_{cm} =
  \displaystyle\frac{1}{A_{2}} \displaystyle\sum_{j\in A_{2}} \mathbf{r}_{j} -
  \displaystyle\frac{1}{A} \displaystyle\sum_{i=1}^{A}\mathbf{r}_{i} = 



  -\sqrt{\displaystyle\frac{A_{1}}{AA_{2}}} \sqrt{\displaystyle\frac{A_{1}A_{2}}{A}}
  \left[ \displaystyle\frac{1}{A_{1}} \displaystyle\sum_{i\in A_{1}}\mathbf{r}_{i} -
  \displaystyle\frac{1}{A_{2}} \displaystyle\sum_{j\in A_{2}}\mathbf{r}_{j} \right].
\end{array}
\label{eq.app.folding.1.9}
\end{equation}
%
Thus
%
\begin{equation}
\begin{array}{lll}
  \mathbf{R}_{1}-\mathbf{R}_{cm} = \sqrt{\displaystyle\frac{A_{2}}{A_{1}A}}\mathbf{q}, &
  \mathbf{R}_{2}-\mathbf{R}_{cm} = -\sqrt{\displaystyle\frac{A_{1}}{A_{2}A}}\mathbf{q},
\end{array}
\label{eq.app.folding.1.10}
\end{equation}
%
where
\begin{equation}
\begin{array}{lll}
  \mathbf{q} =
  \sqrt{\displaystyle\frac{A_{1}A_{2}}{A}}
  \left[ \displaystyle\frac{1}{A_{1}} \displaystyle\sum_{i\in A_{1}} \mathbf{r}_{i} -
  \displaystyle\frac{1}{A_{2}} \displaystyle\sum_{j\in A_{2}} \mathbf{r}_{j} \right]
\end{array}
\label{eq.app.folding.1.11}
\end{equation}
%
is the Jacobi vector determining distance between clusters. Similarly, we
can transform momenta
\begin{equation}
\begin{array}{lll}
  \widehat{\mathbf{P}}_{1}-\widehat{\mathbf{P}}_{cm} =
  \sqrt{\displaystyle\frac{A_{2}}{A_{1}A}}\widehat{\pibf}\mathbf{,} &

  \widehat{\mathbf{P}}_{2}-\widehat{\mathbf{P}}_{cm} =
  -\sqrt{\displaystyle\frac{A_{1}}{A_{2}A}}\widehat{\pibf}\mathbf{,}
\end{array}
\label{eq.app.folding.1.12}
\end{equation}
%
where
\begin{equation}
\begin{array}{lll}
  \widehat{\pibf} =
  \sqrt{\displaystyle\frac{A_{1}A_{2}}{A}}\left[ \displaystyle\frac{1}{A_{1}}
  \displaystyle\sum_{i\in A_{1}}\widehat{\mathbf{p}}_{i} -
  \displaystyle\frac{1}{A_{2}}\displaystyle\sum_{j\in A_{2}}%
  \widehat{\mathbf{p}}_{j}  \right].
\end{array}
\label{eq.app.folding.1.13}
\end{equation}
%
By concluding we can write down
\begin{equation}
\begin{array}{lll}
  \widehat{H}_{e}\left( \mathbf{k}_{\gamma },\varepsilonbf_{\mu}\right)



  & = &
  \exp \left\{ -i \sqrt{\displaystyle\frac{A_{2}}{A_{1}A}}
    \left( \mathbf{k}_{\gamma },\mathbf{q}\right) \right\} \cdot F_{1}^{\left( 1\right) } +
  \displaystyle\frac{e\hbar }{m_{N}c}\sqrt{\displaystyle\frac{A_{2}}{A_{1}A}}
  \exp \left\{ -i\sqrt{\displaystyle\frac{A_{2}}{A_{1}A}}
  \left( \mathbf{k}_{\gamma },\mathbf{q}\right) \right\} \widehat{\pibf}\cdot F_{0}^{\left( 1\right) } \\ 

  & + &
  \exp \left\{ i \sqrt{\displaystyle\frac{A_{1}}{A_{2}A}}
  \left( \mathbf{k}_{\gamma }, \mathbf{q}\right) \right\} \cdot F_{1}^{\left( 2\right) } -
  \displaystyle\frac{e\hbar }{m_{N}c} \sqrt{\displaystyle\frac{A_{1}}{A_{2}A}}
  \exp \left\{ i\sqrt{\displaystyle\frac{A_{1}}{A_{2}A}}\left( \mathbf{k}_{\gamma },\mathbf{q}\right) \right\}
  \widehat{\pibf} \cdot F_{0}^{\left( 2\right) },  
\end{array}
\label{eq.app.folding.1.14}
\end{equation}
%
where
\begin{equation}
\begin{array}{lllllll}
\vspace{1.5mm}
  F_{0}^{\left( \alpha \right) } & = &
  \displaystyle\sum_{i\in A_{\alpha }} \displaystyle\frac{1}{2}\left( 1+ \widehat{\tau }_{iz}\right)
  \exp \left\{ -i\left( \mathbf{k}_{\gamma },
  \mathbf{r}_{i}-\mathbf{R}_{\alpha }\right) \right\}, \\

  F_{1}^{\left( \alpha \right) } & = &
  \displaystyle\frac{e\hbar }{m_{N}c} \displaystyle\sum_{i\in A_{\alpha}}
  \displaystyle\frac{1}{2}\left( 1 + \widehat{\tau }_{iz}\right)
  \exp \left\{ -i\left( \mathbf{k}_{\gamma },\mathbf{r}_{i}-\mathbf{R}_{\alpha }\right) \right\}
  \left( \varepsilonbf_{\mu }\left[ \widehat{\mathbf{p}}_{i}^{\ast } -
  \widehat{\mathbf{P}}_{\alpha }^{\ast }\right] \right)
\end{array}
\label{eq.app.folding.1.15}
\end{equation}
%
with $\alpha =$ 1, 2.

If we consider a two-cluster system in the folding approximation, then a
wave function of the system is
\begin{equation}
  \Psi_{El}=\Phi _{1}\left( A_{1}\right) \Phi _{2}\left( A_{2}\right) R_{El}
  \left( q\right) Y_{lm}\left( \widehat{\mathbf{q}} \right),
\label{eq.app.folding.1.16}
\end{equation}
where functions $\Phi _{1}\left( A_{1}\right) $ and $\Phi _{2}\left(
A_{2}\right) $ describe internal motion of nucleons of the first and second
clusters, respectively, $\widehat{\mathbf{q}}$ \ is a unit vector ($\widehat{%
\mathbf{q}}=\mathbf{q}/q$). Matrix element of the operator
(\ref{eq.app.folding.1.14})
between functions $\Psi _{E_{f} l_{f}}$ and $\Psi_{E_{i} l_{i}}$ is
then equal
\begin{equation}
\begin{array}{lll}
\vspace{1.5mm}
  & &
\left\langle \Psi_{E_{f} l_{f}} \left\vert \widehat{H}_{e}^{\left( 1\right) } \left( \mathbf{k}_{\gamma },
  \varepsilonbf_{\mu }\right) \right\vert \Psi_{E_{i} l_{i}} \right\rangle \\

\vspace{1.5mm}
  & = &
  \left\langle
    R_{E_{f} l_{f}}
  \left( q\right) Y_{l_{f} m_{f}} \left( \widehat{\mathbf{q}}\right) \left\vert
\exp \left\{ -i\sqrt{\displaystyle\frac{A_{2}}{A_{1}A}}\left( \mathbf{k}_{\gamma },%
\mathbf{q}\right) \right\} \right\vert R_{E_{i} l_{i}}
    \left( q\right)
  Y_{l_{i} m_{i}} \left( \widehat{\mathbf{q}}\right) \right\rangle

  \left\langle \Phi_{1} \left( A_{1}\right) \left\vert F_{1}^{\left(
  1\right) }\right\vert \Phi_{1} \left( A_{1}\right) \right\rangle \\

\vspace{1.5mm}
  & + &
  \sqrt{\displaystyle\frac{A_{2}}{A_{1}A}}\left\langle
    R_{E_{f} l_{f}}
    \left( q\right) Y_{l_{f} m_{f}} \left( \widehat{\mathbf{q}}%
\right) \left\vert
  \exp \left\{ -i\sqrt{\displaystyle\frac{A_{2}}{A_{1}A}}
  \left( \mathbf{k}_{\gamma }, \mathbf{q}\right) \right\} \left( \varepsilonbf_{\mu }%
\widehat{\pibf}\right) \right\vert
    R_{E_{i} l_{i}}
    \left( q\right)
    Y_{l_{i} m_{i}}\left( \widehat{\mathbf{q}}\right) \right\rangle

  \left\langle \Phi _{1}\left( A_{1}\right) \left\vert F_{0}^{\left(
  1\right) }\right\vert \Phi _{1}\left( A_{1}\right) \right\rangle \\

\vspace{1.5mm}
  & + &
  \left\langle
    R_{E_{f} l_{f}}
  \left( q\right)
  Y_{l_{f} m_{f}}\left( \widehat{\mathbf{q}}\right) \left\vert
  \exp \left\{ i\sqrt{\displaystyle\frac{A_{1}}{A_{2}A}}\left( \mathbf{k}_{\gamma },
  \mathbf{q}\right) \right\} \right\vert
    R_{E_{i} l_{i}}
  \left( q\right) Y_{l_{i} m_{i}}\left(
  \widehat{\mathbf{q}}\right) \right\rangle

  \left\langle \Phi _{2}\left( A_{2}\right) \left\vert F_{1}^{\left(
  2\right) }\right\vert \Phi _{2}\left( A_{2}\right) \right\rangle \\

\vspace{1.5mm}
  & - &
  \sqrt{\displaystyle\frac{A_{1}}{A_{2}A}}
  \left\langle
    R_{E_{f} l_{f}}
  \left( q\right) Y_{l_{f} m_{f}}\left( \widehat{\mathbf{q}}%
\right) \left\vert \exp \left\{ i\sqrt{\displaystyle\frac{A_{1}}{A_{2}A}}\left( \mathbf{k}%
_{\gamma },\mathbf{q}\right) \right\} \left( \varepsilonbf_{\mu }%
\widehat{\pibf}\right) \right\vert
    R_{E_{i} l_{i}}
  \left( q\right)
  Y_{l_{i} m_{i}} \left( \widehat{\mathbf{q}}\right) \right\rangle

  \left\langle \Phi _{2}\left( A_{2}\right) \left\vert F_{0}^{\left(
  2\right) }\right\vert \Phi _{2}\left( A_{2}\right) \right\rangle.
\end{array}
\label{eq.app.folding.1.17}
\end{equation}
For two $s$-clusters (i.e. for cluster with 1$\leq A_{\alpha }\leq $4 or for
$n$, $p$, $d$, $t$, $^{3}$He and $^{4}$He)

\begin{equation}
\begin{array}{lll}
  \left\langle \Phi _{1}\left( A_{1}\right) \left\vert F_{1}^{\left( 1\right)
}\right\vert \Phi _{1}\left( A_{1}\right) \right\rangle =\left\langle \Phi
_{2}\left( A_{2}\right) \left\vert F_{1}^{\left( 2\right) }\right\vert \Phi
_{2}\left( A_{2}\right) \right\rangle = 0,
\end{array}
\label{eq.app.folding.1.18}
\end{equation}
consequently
\begin{equation}
\begin{array}{lll}
\vspace{1.5mm}
  & & \left\langle \Psi_{E_{f} l_{f}}\left\vert \widehat{H}%
_{e}^{\left( 1\right) }\left( \mathbf{k}_{\gamma },\varepsilonbf%
_{\mu }\right) \right\vert \Psi_{E_{i} l_{i}}\right\rangle \\

\vspace{1.5mm}
  & = &
  \sqrt{\displaystyle\frac{A_{2}}{A_{1}A}}\left\langle
    R_{E_{f} l_{f}}
  \left( q\right)
  Y_{l_{f} m_{f}}\left( \widehat{\mathbf{q}}%
\right) \left\vert \exp \left\{ -i\sqrt{\displaystyle\frac{A_{2}}{A_{1}A}}\left( \mathbf{k%
}_{\gamma },\mathbf{q}\right) \right\} \left( \varepsilonbf_{\mu }%
\widehat{\pibf}\right) \right\vert R_{E_{i} l_{i}}
    \left( q\right)
    Y_{l_{i} m_{i}}\left( \widehat{\mathbf{q}}\right) \right\rangle

  \left\langle \Phi _{1}\left( A_{1}\right) \left\vert F_{0}^{\left(
1\right) }\right\vert \Phi_{1}\, \left( A_{1}\right) \right\rangle \\

  & - &
  \sqrt{\displaystyle\frac{A_{1}}{A_{2}A}}
  \left\langle R_{E_{f} l_{f}}
    \left( q\right) Y_{l_{f} m_{f}}\left( \widehat{\mathbf{q}}%
\right) \left\vert \exp \left\{ i\sqrt{\displaystyle\frac{A_{1}}{A_{2}A}}\left( \mathbf{k}%
_{\gamma },\mathbf{q}\right) \right\} \left( \varepsilonbf_{\mu }%
\widehat{\pibf}\right) \right\vert R_{E_{i} l_{i}}
    \left( q\right) Y_{l_{i} m_{i}} \left( \widehat{\mathbf{q}}\right) \right\rangle 

  \left\langle \Phi _{2}\left( A_{2}\right) \left\vert F_{0}^{\left(
  2\right) }\right\vert \Phi _{2}\left( A_{2}\right) \right\rangle.
\end{array}
\label{eq.app.folding.1.19}
\end{equation}
In the standard approximation of the resonating group method (or cluster
model), form factor
equals
\begin{equation}
\begin{array}{lll}
\left\langle \Phi _{\alpha }\left( A_{\alpha }\right) \left\vert
F_{0}^{\left( 1\right) }\right\vert \Phi _{\alpha }\left( A_{\alpha }\right)
\right\rangle =Z_{\alpha }\exp \left\{ -\displaystyle\frac{1}{4}\displaystyle\frac{A_{\alpha }-1}{%
A_{\alpha }}\left( k_{\gamma }b\right) ^{2}\right\},
\end{array}
\label{eq.app.folding.1.20}
\end{equation}
where $b$ is the oscillator length. Thus, to determin cross section of the
bremsstrahlung emission, we need to calculate matrix element
\begin{equation}
\begin{array}{lll}
  I_{\mu} \left( \alpha \right) =
  \left\langle R_{E_{f} l_{f}} (q)\, Y_{l_{f} m_{f}} (\widehat{\mathbf{q}})
  \left\vert \exp \left\{ - i\alpha \left( \mathbf{k}_{\gamma },\mathbf{q%
}\right) \right\}
  \left( \varepsilonbf_{\mu }\widehat{\pibf} \right) \right\vert
    R_{E_{i} l_{i}}
    \left( q\right) Y_{l_{i} m_{i}} \left( \widehat{%
\mathbf{q}}\right) \right\rangle
\end{array}
\label{eq.app.folding.1.21}
\end{equation}
for two values of the parameter
\begin{equation}
\begin{array}{lll}
  \alpha_{1} = \sqrt{\displaystyle\frac{A_{2}}{A_{1}A}}, \quad
  \alpha_{2} = - \sqrt{\displaystyle\frac{A_{1}}{A_{2}A}}.
\end{array}
\label{eq.app.folding.1.22}
\end{equation}

\section{Multipole expansion of matrix elements
\label{sec.app.multipole}}

\subsection{Matric elements integrated over space coordinates
\label{sec.2.4}}

We shall calculate the following matrix elements
[see Ref.~\cite{Maydanyuk.2012.PRC}, Eqs.~(24)--(41)]:
\begin{equation}
\begin{array}{ll}
  \Bigl< \phi_{f}
    \Bigl| \,  e^{-i\alpha_{i} \mathbf{k_{\gamma}r}} \, \Bigr| \,
      \phi_{i}
      \Bigr>_\mathbf{r} =
  \displaystyle\int
      \phi_{f}^{*}
      (\mathbf{r}) \:
    e^{-i\alpha_{i} \mathbf{k_{\gamma}r}}\:
      \phi_{i}
    (\mathbf{r})\;
    \mathbf{dr}, &

  \hspace{5mm}
  \biggl< \phi_{f}
    \biggl|\,  e^{-i\alpha_{i} \mathbf{k_{\gamma}r}} \displaystyle\frac{\partial}{\partial \mathbf{r}} \,
  \biggr|\, \phi_{i}
  \biggr>_\mathbf{r} =
  \displaystyle\int
      \phi_{f}^{*}
    (\mathbf{r})\:
    e^{-i\alpha_{i} \mathbf{k_{\gamma}r}} \displaystyle\frac{\partial}{\partial \mathbf{r}}\:
    \phi_{i}
    (\mathbf{r})\;
    \mathbf{dr}.
\end{array}
\label{eq.2.4.1.1}
\end{equation}

\subsubsection{Expansion of the vector potential $\mathbf{A}$ by multipoles
\label{sec.2.4.3}}

Let us expand the vectorial potential $\mathbf{A}$ of electromagnetic field by multipole. According to Ref.~\cite{Eisenberg.1973} [see Eqs.~(2.106), p.~58],
in the spherical symmetric approximation we have:
\begin{equation}
  \xibf_{\mu}\, e^{i \alpha_{i} \mathbf{k_{\gamma}r}} =
    \mu\, \sqrt{2\pi}\, \sum_{l_{\gamma}=1}\,
    (2l_{\gamma}+1)^{1/2}\, i^{l_{\gamma}}\,  \cdot
    \Bigl[ \mathbf{A}_{l_{\gamma}\mu} (\mathbf{r}, M) +
    i\mu\, \mathbf{A}_{l_{\gamma}\mu} (\mathbf{r}, E) \Bigr],
\label{eq.2.4.3.1}
\end{equation}
where [see~\cite{Eisenberg.1973}, (2.73) in p.~49, (2.80) in p.~51]
\begin{equation}
\begin{array}{lcl}
  \vspace{2mm}
  \mathbf{A}_{l_{\gamma}\mu}(\mathbf{r}, M) & = &
        j_{l_{\gamma}}(\alpha_{i} k_{\gamma}r) \: \mathbf{T}_{l_{\gamma}l_{\gamma},\mu} (\widehat{\mathbf{r}}), \\
  \vspace{2mm}
  \mathbf{A}_{l_{\gamma}\mu}(\mathbf{r}, E) & = &
        \sqrt{\displaystyle\frac{l_{\gamma}+1}{2l_{\gamma}+1}}\,
        j_{l_{\gamma}-1}(\alpha_{i} k_{\gamma}r) \: \mathbf{T}_{l_{\gamma}l_{\gamma}-1,\mu}(\widehat{\mathbf{r}})\; -
        \sqrt{\displaystyle\frac{l_{\gamma}}{2l_{\gamma}+1}}\,
        j_{l_{\gamma}+1}(\alpha_{i} k_{\gamma}r) \: \mathbf{T}_{l_{\gamma}l_{\gamma}+1,\mu}(\widehat{\mathbf{r}}).
\end{array}
\label{eq.2.4.3.2}
\end{equation}
Here, $\mathbf{A}_{l_{\gamma}\mu}(\textbf{r}, M)$ and $\mathbf{A}_{l_{\gamma}\mu}(\textbf{r}, E)$ are \emph{magnetic} and \emph{electric multipoles},
$j_{l_{\gamma}}(\alpha_{i} k_{\gamma}r)$ is \emph{spherical Bessel function of order $l_{\gamma}$},
$\mathbf{T}_{l_{\gamma}l_{\gamma}^{\prime},\mu}(\widehat{\mathbf{r}})$ are \emph{vector spherical harmonics},
 $\xibf_{\mu}$ are \emph{vectors of circular polarization} of emitted photon.
Eq.~(\ref{eq.2.4.3.1}) is solution of the wave equation of electromagnetic field in form of plane wave, which is presented as summation of the electrical and magnetic multipoles (for example, see p.~83--92 in~\cite{Ahiezer.1981}). Therefore, separate multipolar terms in eq.~(\ref{eq.2.4.3.1}) are solutions of this wave equation for chosen numbers $j_{\gamma}$ and $l_{\gamma}$ ($j_{\gamma}$ is quantum number characterizing eigenvalue of the full momentum operator, while $l_{\gamma}= j_{\gamma}-1, j_{\gamma}, j_{\gamma}+1$ is connected with orbital momentum operator, but it defines eigenvalues of photon parity and, so, it is quantum number also).


We orient the frame so that axis $z$ be directed along the vector $\mathbf{k}_{\gamma}$ (see~\cite{Eisenberg.1973}, (2.105) in p.~57). According to \cite{Eisenberg.1973} (see p.~45), the functions $\mathbf{T}_{l_{\gamma}l_{\gamma}^{\prime},\mu}(\widehat{\mathbf{r}})$ have the following form
($\xibf_{0} = 0$):
\begin{equation}
  \mathbf{T}_{j_{\gamma}l_{\gamma},m} (\widehat{\mathbf{r}}) =
  \sum\limits_{\mu = \pm 1} (l_{\gamma}, 1, j_{\gamma} \,\big| \,m-\mu, \mu, m)\;
  Y_{l_{\gamma},m-\mu}(\widehat{\mathbf{r}})\;
  \xibf_{\mu},
\label{eq.2.4.3.3}
\end{equation}
where $(l, 1, j \,\bigl| \, m-\mu, \mu, m)$ are \emph{Clebsh-Gordon coefficients},
$Y_{lm}(\theta, \varphi)$ are \emph{spherical functions} defined, according to~\cite{Landau.v3.1989} (see p.~119, (28,7)--(28,8)).
From eq.~(\ref{eq.2.4.3.1}) one can obtain such a formula (at $\varepsilonbf^{(3)}=0$):
\begin{equation}
  e^{-i \alpha_{i} \mathbf{k_{\gamma}r}} =
  \displaystyle\frac{1}{2}\,
  \displaystyle\sum\limits_{\mu = \pm 1}
    \xibf_{\mu}\, \mu\, \sqrt{2\pi}\, \sum_{l_{\gamma}=1}\,
    (2l_{\gamma}+1)^{1/2}\, (-i)^{l_{\gamma}}\,  \cdot
    \Bigl[ \mathbf{A}_{l_{\gamma}\mu}^{*} (\mathbf{r}, M) -
    i\mu\, \mathbf{A}_{l_{\gamma}\mu}^{*} (\mathbf{r}, E) \Bigr].
\label{eq.2.4.3.5}
\end{equation}

\subsubsection{Central interactions
\label{sec.2.4.4}}

Using (\ref{eq.2.4.3.5}), for (\ref{eq.2.4.1.1}) we find:
\begin{equation}
\begin{array}{ll}
  \vspace{1mm}
  \Bigl< \phi_{f}
  \Bigl| \,  e^{-i\alpha_{i} \mathbf{k_{\gamma}r}} \, \Bigr|\, \phi_{i}
  \Bigr>_\mathbf{r} =
  \sqrt{\displaystyle\frac{\pi}{2}}\:
  \displaystyle\sum\limits_{l_{\gamma}=1}\,
    (-i)^{l_{\gamma}}\, \sqrt{2l_{\gamma}+1}\;
  \displaystyle\sum\limits_{\mu = \pm 1}
    \Bigl[ \mu\,\tilde{p}_{l_{\gamma}\mu}^{M} - i\, \tilde{p}_{l_{\gamma}\mu}^{E} \Bigr], \\

  \biggl< \phi_{f}
  \biggl| \,  e^{-i\alpha_{i} \mathbf{k_{\gamma}r}} \displaystyle\frac{\partial}{\partial \mathbf{r}}\,
  \biggr|\, \phi_{i}
  \biggr>_\mathbf{r} =
  \sqrt{\displaystyle\frac{\pi}{2}}\:
  \displaystyle\sum\limits_{l_{\gamma}=1}\,
    (-i)^{l_{\gamma}}\, \sqrt{2l_{\gamma}+1}\;
  \displaystyle\sum\limits_{\mu = \pm 1}
    \xibf_{\mu}\, \mu\, \times
    \Bigl[ p_{l_{\gamma}\mu}^{M} - i\mu\: p_{l_{\gamma}\mu}^{E} \Bigr],
\end{array}
\label{eq.2.4.4.1}
\end{equation}
where
\begin{equation}
\begin{array}{lcllcl}
  p_{l_{\gamma}\mu}^{M} & = &
    \displaystyle\int
        \phi^{*}_{f}(\mathbf{r}) \,
        \biggl( \displaystyle\frac{\partial}{\partial \mathbf{r}}\, \phi_{i}(\mathbf{r}) \biggr) \,
        \mathbf{A}_{l_{\gamma}\mu}^{*} (\mathbf{r}, M) \;
        \mathbf{dr}, &

  \hspace{7mm}
  p_{l_{\gamma}\mu}^{E} & = &
    \displaystyle\int
        \phi^{*}_{f}(\mathbf{r}) \,
        \biggl( \displaystyle\frac{\partial}{\partial \mathbf{r}}\, \phi_{i}(\mathbf{r}) \biggr)\,
        \mathbf{A}_{l_{\gamma}\mu}^{*} (\mathbf{r}, E) \;
        \mathbf{dr},
\end{array}
\label{eq.2.4.4.2}
\end{equation}
and
\begin{equation}
\begin{array}{lcllcl}
  \tilde{p}_{l_{\gamma} \mu}^{M} & = &
    \xibf_{\mu}\,
    \displaystyle\int
      \phi^{*}_{f}(\mathbf{r})\,
      \phi_{i}(\mathbf{r})\;
      \mathbf{A}_{l_{\gamma}\mu}^{*} (\mathbf{r}, M) \;
      \mathbf{dr}, &
  \hspace{7mm}
  \tilde{p}_{l_{\gamma}\mu}^{E} & = &
    \xibf_{\mu}\,
    \displaystyle\int
      \phi^{*}_{f}(\mathbf{r})\,
      \phi_{i}(\mathbf{r})\;
      \mathbf{A}_{l_{\gamma}\mu}^{*} (\mathbf{r}, E)\;
      \mathbf{dr}.
\end{array}
\label{eq.2.4.4.3}
\end{equation}

\subsubsection{Calculations of the components $p_{l_{\gamma}\mu}^{M}$, $p_{l_{\gamma}\mu}^{E}$ and\,
$\tilde{p}_{l_{\gamma}\mu}^{M}$, $\tilde{p}_{l_{\gamma}\mu}^{E}$
\label{sec.2.4.6}}

For calculation of these components we shall use \emph{gradient formula} (see~\cite{Eisenberg.1973}, (2.56) in p.~46):
\begin{equation}
\begin{array}{lcl}
  \displaystyle\frac{\partial}{\partial \mathbf{r}}\: \phi_{i}
  (\mathbf{r}) & = &
  \displaystyle\frac{\partial}{\partial \mathbf{r}}\:
    \Bigl\{ R_{i} (r)\: Y_{l_{i}m_{i}}(\widehat{\mathbf{r}}) \Bigr\} =

    \sqrt{\displaystyle\frac{l_{i}}{2l_{i}+1}}\:
    \biggl( \displaystyle\frac{dR_{i}(r)}{dr} + \displaystyle\frac{l_{i}+1}{r}\, R_{i}(r) \biggr)\,
      \mathbf{T}_{l_{i} l_{i}-1, m_{i}}(\widehat{\mathbf{r}}) - \\
  & - &
  \sqrt{\displaystyle\frac{l_{i}+1}{2l_{i}+1}}\:
    \biggl( \displaystyle\frac{dR_{i}(r)}{dr} - \displaystyle\frac{l_{i}}{r}\, R_{i}(r) \biggr)\,
      \mathbf{T}_{l_{i} l_{i}+1, m_{i}}(\widehat{\mathbf{r}}),
\end{array}
\label{eq.2.4.6.1}
\end{equation}
and obtain:
\begin{equation}
\begin{array}{lcl}
\vspace{1mm}
  p_{l_{\rm ph,\mu}}^{M} & = &
    \sqrt{\displaystyle\frac{l_{i}}{2l_{i}+1}}\:
      I_{M}(l_{i},l_{f}, l_{\gamma}, l_{i}-1, \mu) \cdot
      \Bigl\{
        J_{1}(l_{i},l_{f},l_{\gamma}, \alpha_{i}) + (l_{i}+1) \cdot J_{2}(l_{i},l_{f},l_{\gamma}, \alpha_{i})
      \Bigr\}\; - \\
\vspace{3mm}
  & - &
    \sqrt{\displaystyle\frac{l_{i}+1}{2l_{i}+1}}\:
      I_{M}(l_{i},l_{f}, l_{\gamma}, l_{i}+1, \mu) \cdot
      \Bigl\{
        J_{1}(l_{i},l_{f},l_{\gamma}, \alpha_{i}) - l_{i} \cdot J_{2}(l_{i},l_{f},l_{\gamma}, \alpha_{i})
      \Bigr\}, \\

\vspace{1mm}
  p_{l_{\rm ph,\mu}}^{E} & = &
    \sqrt{\displaystyle\frac{l_{i}\,(l_{\gamma}+1)}{(2l_{i}+1)(2l_{\gamma}+1)}} \cdot
      I_{E}(l_{i},l_{f}, l_{\gamma}, l_{i}-1, l_{\gamma}-1, \mu) \cdot
      \Bigl\{
        J_{1}(l_{i},l_{f},l_{\gamma}-1, \alpha_{i})\; +
        (l_{i}+1) \cdot J_{2}(l_{i},l_{f},l_{\gamma}-1, \alpha_{i})
      \Bigr\}\; - \\
\vspace{1mm}
    & - &
    \sqrt{\displaystyle\frac{l_{i}\,l_{\gamma}}{(2l_{i}+1)(2l_{\gamma}+1)}} \cdot
      I_{E} (l_{i},l_{f}, l_{\gamma}, l_{i}-1, l_{\gamma}+1, \mu) \cdot
      \Bigl\{
        J_{1}(l_{i},l_{f},l_{\gamma}+1, \alpha_{i})\; +
        (l_{i}+1) \cdot J_{2}(l_{i},l_{f},l_{\gamma}+1, \alpha_{i})
      \Bigr\}\; + \\
\vspace{1mm}
  & + &
    \sqrt{\displaystyle\frac{(l_{i}+1)(l_{\gamma}+1)}{(2l_{i}+1)(2l_{\gamma}+1)}} \cdot
      I_{E} (l_{i},l_{f},l_{\gamma}, l_{i}+1, l_{\gamma}-1, \mu) \cdot
      \Bigl\{
        J_{1}(l_{i},l_{f},l_{\gamma}-1, \alpha_{i})\; -
        l_{i} \cdot J_{2}(l_{i},l_{f},l_{\gamma}-1, \alpha_{i})
      \Bigr\}\; - \\
  & - &
    \sqrt{\displaystyle\frac{(l_{i}+1)\,l_{\gamma}}{(2l_{i}+1)(2l_{\gamma}+1)}} \cdot
      I_{E} (l_{i},l_{f}, l_{\gamma}, l_{i}+1, l_{\gamma}+1, \mu) \cdot
      \Bigl\{
        J_{1}(l_{i},l_{f},l_{\gamma}+1, \alpha_{i})\; -
        l_{i} \cdot J_{2}(l_{i},l_{f},l_{\gamma}+1, \alpha_{i})
      \Bigr\},
\end{array}
\label{eq.2.4.6.4}
\end{equation}
where
\begin{equation}
\begin{array}{ccl}
  J_{1}(l_{i},l_{f},n, \alpha_{i}) & = &
  \displaystyle\int\limits^{+\infty}_{0}
    \displaystyle\frac{dR_{i}(r, l_{i})}{dr}\: R^{*}_{f}(l_{f},r)\,
    j_{n}(\alpha_{i}\, k_{\gamma}r)\; r^{2} dr, \\

  J_{2}(l_{i},l_{f},n, \alpha_{i}) & = &
  \displaystyle\int\limits^{+\infty}_{0}
    R_{i}(r, l_{i})\, R^{*}_{f}(l_{f},r)\: j_{n}(\alpha_{i}\, k_{\gamma}r)\; r\, dr, \\

  I_{M}\, (l_{i}, l_{f}, l_{\gamma}, l_{1}, \mu) & = &
    \displaystyle\int
      Y_{l_{f}m_{f}}^{*}(\widehat{\mathbf{r}})\,
      \mathbf{T}_{l_{i}\, l_{1},\, m_{i}}(\widehat{\mathbf{r}})\,
      \mathbf{T}_{l_{\gamma}\,l_{\gamma},\, \mu}^{*}(\widehat{\mathbf{r}})\; d\Omega, \\

  I_{E}\, (l_{i}, l_{f}, l_{\gamma}, l_{1}, l_{2}, \mu) & = &
    \displaystyle\int
      Y_{l_{f}m_{f}}^{*}(\widehat{\mathbf{r}})\,
      \mathbf{T}_{l_{i} l_{1},\, m_{i}}(\widehat{\mathbf{r}})\,
      \mathbf{T}_{l_{\gamma} l_{2},\, \mu}^{*}(\widehat{\mathbf{r}})\; d\Omega.
\end{array}
\label{eq.2.4.6.3}
\end{equation}
By the same way for $\tilde{p}_{l_{\gamma}\mu}^{M}$ and $\tilde{p}_{l_{\gamma}\mu}^{E}$ we find:
%
\begin{equation}
\begin{array}{lcl}
  \tilde{p}_{l_{\gamma}\mu}^{M} & = &
    \tilde{I}\,(l_{i},l_{f},l_{\gamma}, l_{\gamma}, \mu) \cdot \tilde{J}\, (l_{i},l_{f},l_{\gamma}, \alpha_{i}), \\
  \tilde{p}_{l_{\gamma}\mu}^{E} & = &
    \sqrt{\displaystyle\frac{l_{\gamma}+1}{2l_{\gamma}+1}}
      \tilde{I}\,(l_{i},l_{f},l_{\gamma},l_{\gamma}-1,\mu) \cdot \tilde{J}\,(l_{i},l_{f},l_{\gamma}-1, \alpha_{i}) -
    \sqrt{\displaystyle\frac{l_{\gamma}}{2l_{\gamma}+1}}
      \tilde{I}\,(l_{i},l_{f},l_{\gamma},l_{\gamma}+1,\mu) \cdot \tilde{J}\,(l_{i},l_{f},l_{\gamma}+1, \alpha_{i}),
\end{array}
\label{eq.2.4.6.7}
\end{equation}
where
\begin{equation}
\begin{array}{lcl}
  \tilde{J}\,(l_{i},l_{f},n, \alpha_{i}) & = &
  \displaystyle\int\limits^{+\infty}_{0}
    R_{i}(r)\, R^{*}_{f}(l,r)\, j_{n}(\alpha_{i}\, k_{\gamma}r)\; r^{2} dr, \\

  \tilde{I}\,(l_{i}, l_{f}, l_{\gamma}, n, \mu) & = &
  \xibf_{\mu} \displaystyle\int
    Y_{l_{i}m_{i}}(\widehat{\mathbf{r}})\:
    Y_{l_{f}m_{f}}^{*}(\widehat{\mathbf{r}})\:
    \mathbf{T}_{l_{\gamma} n,\mu}^{*}(\widehat{\mathbf{r}}) \: d\Omega.
\end{array}
\label{eq.2.4.6.6}
\end{equation}

\section{Linear and circular polarizations of the photon emitted
\label{sec.app.polarization}}

We define \emph{vectors of linear polarization} of emitted photon as (in Coulomb gauge at $\varepsilonbf^{(3)}=0$):
\begin{equation}
\begin{array}{lll}
  \varepsilonbf^{(1)} = \displaystyle\frac{1}{\sqrt{2}}\: \bigl( \xibf_{-1} - \xibf_{+1} \bigr),
  \hspace{3mm}
  \varepsilonbf^{(2)} = \displaystyle\frac{i}{\sqrt{2}}\: \bigl( \xibf_{-1} + \xibf_{+1} \bigr),
\end{array}
\label{eq.app.polarization.1}
\end{equation}
where
$\xibf_{\mu}$ are \emph{vectors of circular polarization} with opposite directions of rotation for the emitted photon used in Eqs.~(\ref{eq.2.4.3.1}).
Rewrite vectors of linear polarization $\varepsilonbf^{(\alpha)}$ through vectors of circular polarization $\xibf_{\mu}$ (see Ref.~\cite{Eisenberg.1973}, (2.39), p.~42;
Appendix~A in Ref.~\cite{Maydanyuk.2012.PRC},
$\varepsilonbf^{(\alpha), *} = \varepsilonbf^{(\alpha)}$):
\begin{equation}
\begin{array}{lll}
  \xibf_{-1} = \displaystyle\frac{1}{\sqrt{2}} (\varepsilonbf^{(1)} - i\varepsilonbf^{(2)}), &
  \xibf_{+1} = -\displaystyle\frac{1}{\sqrt{2}} (\varepsilonbf^{(1)} + i\varepsilonbf^{(2)}), &
  \xibf_{0} = \varepsilonbf^{(3)}.
\end{array}
\label{eq.app.polarization.2}
\end{equation}
We obtain properties:
\begin{equation}
\begin{array}{ccc}
  \displaystyle\sum\limits_{\alpha = 1,2} \varepsilonbf^{(\alpha),*} =
    h_{-1} \xibf_{-1}^{*} + h_{+1} \xibf_{+1}^{*},
\end{array}
\label{eq.app.polarization.3}
\end{equation}
\begin{equation}
\begin{array}{ccc}
  \displaystyle\sum\limits_{\mu = \pm 1} \xibf_{\mu}^{*} \cdot \xibf_{\mu} =
  \displaystyle\frac{1}{2}\,
    \bigl(\varepsilonbf^{(1)} - i\varepsilonbf^{(2)}\bigr)\, \bigl(\varepsilonbf^{(1)} - i\varepsilonbf^{(2)}\bigr)^{*} +
  \displaystyle\frac{1}{2}\,
    \bigl(\varepsilonbf^{(1)} + i\varepsilonbf^{(2)}\bigr)\, \bigl(\varepsilonbf^{(1)} + i\varepsilonbf^{(2)}\bigr)^{*} = 2,
\end{array}
\label{eq.app.polarization.4}
\end{equation}
where
\begin{equation}
\begin{array}{lcr}
  h_{-1} = \displaystyle\frac{1}{\sqrt{2}} (1-i), &
  h_{1}  = - \displaystyle\frac{1}{\sqrt{2}} (1+i), &
  h_{-1} + h_{1} = -i \sqrt{2}.
\end{array}
\label{eq.app.polarization.5}
\end{equation}
Also there is property (see Eqs.~(3), (5) in Ref.~\cite{Maydanyuk.2006.EPJA}):
\begin{equation}
\begin{array}{lcr}
  \displaystyle\sum_{\alpha = 1,2} \varepsilonbf^{\rm (\alpha),\, *} =
  \displaystyle\sum\limits_{\mu = \pm 1} h_{m}\, \xibf_{m}^{*}.
\end{array}
\label{eq.app.polarization.6}
\end{equation}

We shall find multiplications of vectors $\xibf_{\pm 1}$. From Eq.~(\ref{eq.app.polarization.2}) we obtain:
\begin{equation}
\begin{array}{ll}
  \xibf_{-1}^{*} =
  \Bigl( \displaystyle\frac{1}{\sqrt{2}} (\varepsilonbf^{(1)} - i\varepsilonbf^{(2)}) \Bigr)^{*} =
  \displaystyle\frac{1}{\sqrt{2}} (\varepsilonbf^{(1)} + i\varepsilonbf^{(2)}) =
  -\, \xibf_{+1}, &
  \xibf_{+1}^{*} = -\, \xibf_{-1}
\end{array}
\label{eq.app.polarization.7}
\end{equation}
and
\begin{equation}
\begin{array}{cc}
  \xibf_{-1}^{*} = - \xibf_{+1}, &
  \xibf_{+1}^{*} = - \xibf_{-1}.
\end{array}
\label{eq.app.polarization.8}
\end{equation}
We check orthogonality conditions as
\begin{equation}
\begin{array}{llllll}
\vspace{1.5mm}
  \xibf_{-1} \cdot \xibf_{-1} = \xibf_{+1} \cdot \xibf_{+1} = 0, &
  \hspace{5mm}
  \xibf_{-1} \cdot \xibf_{-1}^{*} = \xibf_{+1} \cdot \xibf_{+1}^{*} = -1, \\

  \xibf_{-1} \cdot \xibf_{+1} = 1, &
  \hspace{5mm}
  \xibf_{-1} \cdot \xibf_{+1}^{*} = \xibf_{+1} \cdot \xibf_{-1}^{*} = 0.
\end{array}
\label{eq.app.polarization.9}
\end{equation}
We calculate multiplications of vectors as
\begin{equation}
\begin{array}{lcllcl}
  \vspace{2mm}
  \Bigl[\xibf_{-1}^{*} \times \xibf_{+1}\Bigr] & = & -\, \Bigl[\xibf_{+1} \times \xibf_{+1}\Bigr] = 0, &

  \hspace{10mm}
  \Bigl[\xibf_{-1}^{*} \times \xibf_{-1}\Bigr] & = & -\, \Bigl[\xibf_{+1} \times \xibf_{-1}\Bigr], \\

  \Bigl[\xibf_{+1}^{*} \times \xibf_{-1}\Bigr] & = & -\, \Bigl[\xibf_{-1} \times \xibf_{-1}\Bigr] = 0, &

  \hspace{10mm}
  \Bigl[\xibf_{+1}^{*} \times \xibf_{+1}\Bigr] & = & -\, \Bigl[\xibf_{-1} \times \xibf_{+1}\Bigr]
\end{array}
\label{eq.app.polarization.10}
\end{equation}
and
\begin{equation}
\begin{array}{lll}

  \Bigl[ \xibf_{-1} \times \xibf_{-1}^{*} \Bigr] & = &
  -\, \Bigl[ \xibf_{+1} \times \xibf_{+1}^{*} \Bigr] =
  i\, \bigl[ \varepsilonbf^{(1)} \times \varepsilonbf^{(2)} \bigr] =
  -\, \bigl[ \xibf_{-1} \times \xibf_{+1} \bigr].
\end{array}
\label{eq.app.polarization.11}
\end{equation}

Now we take into account that two vectors $\varepsilonbf^{(1)}$ and $\varepsilonbf^{(2)}$ are vectors of polarization of photon emitted, which are perpendicular to direction of emission of this photon defined by vector $\vb{k}$. Modulus of vectorial multiplication $\bigl[ \varepsilonbf^{(1)} \times \varepsilonbf^{(2)} \bigr]$ equals to unity.
So, we have:
\begin{equation}
\begin{array}{lll}
  \bigl[ \varepsilonbf^{(1)} \times \varepsilonbf^{(2)} \bigr] =
  \displaystyle\frac{\vb{k}_{\gamma}}{\bigl| \vb{k}_{\gamma} \bigr|} \equiv
  \widehat{\gamma}.
\end{array}
\label{eq.app.polarization.12}
\end{equation}
By such a basis, we rewrite properties (\ref{eq.app.polarization.11}) as
\begin{equation}
\begin{array}{lll}
  \Bigl[ \xibf_{-1} \times \xibf_{-1}^{*} \Bigr] & = &
  -\, \Bigl[ \xibf_{+1} \times \xibf_{+1}^{*} \Bigr] =
  -\, \bigl[ \xibf_{-1} \times \xibf_{+1} \bigr] =
  i\, \bigl[ \varepsilonbf^{1} \times \varepsilonbf^{2} \bigr] =
  i\, \widehat{\mathbf{\gamma}}.
\end{array}
\label{eq.app.polarization.13}
\end{equation}


\section{Angular integrals $I_{E}$, $I_{M}$ and $\tilde{I}$
\label{sec.app.2}}

We shall calculate the integrals in eqs.~(\ref{eq.2.4.6.3}) and (\ref{eq.2.4.6.6})
[see Appendix~B in Ref.~\cite{Maydanyuk.2012.PRC}]:
\begin{equation}
\begin{array}{lcl}
  I_{M}\, (l_{i}, l_{f}, l_{\gamma}, l_{1}, \mu) & = &
    \displaystyle\int
      Y_{l_{f}m_{f}}^{*}(\widehat{\mathbf{r}})\,
      \mathbf{T}_{l_{i}\, l_{1},\, m_{i}}(\widehat{\mathbf{r}})\,
      \mathbf{T}_{l_{\gamma}\,l_{\gamma},\, \mu}^{*}(\widehat{\mathbf{r}})\; d\Omega, \\

  I_{E}\, (l_{i}, l_{f}, l_{\gamma}, l_{1}, l_{2}, \mu) & = &
    \displaystyle\int
      Y_{l_{f}m_{f}}^{*}(\widehat{\mathbf{r}})\,
      \mathbf{T}_{l_{i} l_{1},\, m_{i}}(\widehat{\mathbf{r}})\,
      \mathbf{T}_{l_{\gamma} l_{2},\, \mu}^{*}(\widehat{\mathbf{r}})\; d\Omega, \\

  \tilde{I}\, (l_{i}, l_{f}, l_{\gamma}, n, \mu) & = &
    \xibf_{\mu} \displaystyle\int
      Y_{l_{f}m_{f}}^{*}(\widehat{\mathbf{r}})\,
      Y_{l_{i}m_{i}}(\widehat{\mathbf{r}})\,
      \mathbf{T}_{l_{\gamma} n,\, \mu}^{*}(\widehat{\mathbf{r}})\; d\Omega.
\end{array}
\label{eq.app.2.1}
\end{equation}
Substituting the function $\mathbf{T}_{jl,m}(\widehat{\mathbf{r}})$ defined by eq.~(\ref{eq.2.4.3.3}),
we obtain (at ${\mathbf \xi}_{0} = 0$):
\begin{equation}
\begin{array}{lcl}
  I_{M}\, (l_{i}, l_{f}, l_{\gamma}, l_{1}, \mu) & = &
    \displaystyle\sum\limits_{\mu^{\prime} = \pm 1}
      (l_{1}, 1, l_{i} \,\big| \,m_{i}-\mu^{\prime}, \mu^{\prime}, m_{i})\;
      (l_{\gamma}, 1, l_{\gamma} \,\big|\, \mu-\mu^{\prime}, \mu^{\prime}, \mu)\; \times \\
  & \times &
    \displaystyle\int
      Y_{l_{f}m}^{*}(\widehat{\mathbf{r}}) \cdot
      Y_{l_{1},\, m_{i}-\mu^{\prime}}(\widehat{\mathbf{r}}) \cdot
      Y_{l_{\gamma},\, \mu-\mu^{\prime}}^{*} (\widehat{\mathbf{r}})\; d\Omega, \\

  I_{E}\, (l_{i}, l_{f}, l_{\gamma}, l_{1}, l_{2}, \mu) & = &
    \displaystyle\sum\limits_{\mu^{\prime} = \pm 1}
      (l_{1}, 1, l_{i} \,\big| \,m_{i}-\mu^{\prime}, \mu^{\prime}, m_{i})\;
      (l_{2}, 1, l_{\gamma} \,\big|\, \mu-\mu^{\prime}, \mu^{\prime}, \mu)\; \times \\
  & \times &
    \displaystyle\int
      Y_{l_{f}m}^{*}(\widehat{\mathbf{r}}) \cdot
      Y_{l_{1},\, m_{i}-\mu^{\prime}} (\widehat{\mathbf{r}}) \cdot
      Y_{l_{2},\, \mu-\mu^{\prime}}^{*} (\widehat{\mathbf{r}})\; d\Omega.
\end{array}
\label{eq.app.2.2}
\end{equation}
\begin{equation}
\begin{array}{ccl}
  \tilde{I}\, (l_{i}, l_{f}, l_{\gamma}, n, \mu) & = &
    (n, 1, l_{\gamma} \,\big| \,0, \mu, \mu) \times
    \displaystyle\int
      Y_{l_{f}m_{f}}^{*}(\widehat{\mathbf{r}})\,
      Y_{l_{i}m_{i}}(\widehat{\mathbf{r}})\,
      Y_{n0}^{*}(\widehat{\mathbf{r}})\; d\Omega.
\end{array}
\label{eq.app.2.3}
\end{equation}
Here, we have taken orthogonality of vectors $\xi_{\pm 1}$ into account.
In these formulas we shall find angular integral:
\begin{equation}
\begin{array}{l}
\vspace{1mm}
  \displaystyle\int \:
    Y_{l_{f}m_{f}}^{*}(\widehat{\mathbf{r}})\,
    Y_{l_{1},\, m_{i}-\mu^{\prime}}(\widehat{\mathbf{r}})\,
    Y_{n,\, \mu-\mu^{\prime}}^{*}(\widehat{\mathbf{r}})\;
    d\Omega =

    (-1)^{l_{f} + n + m_{i} - \mu^{\prime}}\;
    i^{l_{f}+l_{1}+n + |m_{f}| + |m_{i} - \mu^{\prime}| + |m_{i} - m_{f}-\mu^{\prime}|}\; \times \\

  \;\times
    \sqrt{
      \displaystyle\frac{(2l_{f}+1)\, (2l_{1}+1)\, (2n+1)}{16\pi}
      \displaystyle\frac{(l_{f}-|m_{f}|)!}{(l_{f}+|m_{f}|)!}\;
      \displaystyle\frac{(l_{1}-|m_{i}-\mu^{\prime}|)!} {(l_{1}+|m_{i}-\mu^{\prime}|)!}\;
      \displaystyle\frac{(n-|m_{i} - m_{f}-\mu^{\prime}|)!}{(n+|m_{i} - m_{f} -\mu^{\prime}|)!}}\;
    \times \\

  \;\times
    \displaystyle\int\limits_{0}^{\pi}\:
      P_{l_{f}}^{|m_{f}|}(\cos{\theta})\;
      P_{l_{1}}^{|m_{i} - \mu^{\prime}|}(\cos{\theta})\;
      P_{n}^{|m_{i} - m_{f} - \mu^{\prime}|} (\cos{\theta}) \cdot
      \sin{\theta}\, d\theta,
\end{array}
\label{eq.app.2.4}
\end{equation}
where $P_{l}^{m}(\cos{\theta})$ are associated Legandre's polynomials, and we obtain conditions:
\begin{equation}
\begin{array}{llllll}
  \mbox{for integrals } I_{M}, I_{E}: &
  \mu = m_{i} - m_{f}, &
  n \ge |\mu - \mu^{\prime}| = |m_{i} - m_{f} + \mu^{\prime}|, & \mu = \pm 1, \\

  \mbox{for integral } \tilde{I}: &
  m_{i} = m_{f}. & &
\end{array}
\label{eq.app.2.5}
\end{equation}
Using formula (\ref{eq.app.2.4}), we calculate integrals (\ref{eq.app.2.2}) and (\ref{eq.app.2.3}):
\begin{equation}
\begin{array}{lcl}
  I_{M}\, (l_{i}, l_{f}, l_{\gamma}, l_{1}, \mu) & = &
    \delta_{\mu, m_{i}-m_{f}}\;
    \displaystyle\sum\limits_{\mu^{\prime} = \pm 1}
      C_{l_{i} l_{f} l_{ph} l_{1} l_{ph}}^{m_{i} m_{f} \mu^{\prime}}
      \displaystyle\int\limits_{0}^{\pi}\:
        f_{l_{1} l_{f} l_{\gamma}}^{m_{i} m_{f} \mu^{\prime}}(\theta)\; \sin{\theta}\,d\theta, \\

  I_{E}\, (l_{i}, l_{f}, l_{\gamma}, l_{1}, l_{2}, \mu) & = &
    \delta_{\mu, m_{i}-m_{f}}\;
    \displaystyle\sum\limits_{\mu^{\prime} = \pm 1}
      C_{l_{i} l_{f} l_{ph} l_{1} l_{2}}^{m_{i} m_{f} \mu^{\prime}}
      \displaystyle\int\limits_{0}^{\pi} \:
      f_{l_{1} l_{f} l_{2}}^{m_{i} m_{f} \mu^{\prime}}(\theta)\; \sin{\theta}\,d\theta, \\

  \tilde{I}\, (l_{i}, l_{f}, l_{\gamma}, n, \mu) & = &
    C_{l_{i} l_{f} l_{\gamma} n}^{m_{i} \mu}
    \displaystyle\int\limits_{0}^{\pi}\:
    f_{l_{i} l_{f} n}^{m_{i} m_{i} 0}(\theta)\; \sin{\theta}\,d\theta,
\end{array}
\label{eq.app.2.6}
\end{equation}

\vspace{-5mm}
\noindent
where
\begin{equation}
\begin{array}{lcl}
\vspace{1mm}
  C_{l_{i} l_{f} l_{ph} l_{1} l_{2}}^{m_{i} m_{f} \mu^{\prime}} & = &
    (-1)^{l_{f} + l_{2} + m_{i} - \mu^{\prime}}\;
    i^{l_{f}+l_{1}+ l_{2} + |m_{f}| + |m_{i} - \mu^{\prime}| + |m_{i} - m_{f}-\mu^{\prime}|}\; \times \\
  \vspace{2mm}
  & \times &
    (l_{1}, 1, l_{i} \,\big| \,m_{i}-\mu^{\prime}, \mu^{\prime}, m_{i})\;
    (l_{2}, 1, l_{\gamma} \,\big|\, m_{i}-m_{f} -\mu^{\prime}, \mu^{\prime}, m_{i}-m_{f})\; \times \\
  & \times &
    \sqrt{
      \displaystyle\frac{(2l_{f}+1)\, (2l_{1}+1)\, (2l_{2}+1)}{16\pi}
      \displaystyle\frac{(l_{f}-|m_{f}|)!}{(l_{f}+|m_{f}|)!}\;
      \displaystyle\frac{(l_{1}-|m_{i}-\mu^{\prime}|)!}{(l_{1}+|m_{i}-\mu^{\prime}|)!}\;
      \displaystyle\frac{(l_{2}-|m_{i}-m_{f}-\mu^{\prime}|)!} {(l_{2} +|m_{i} - m_{f} -\mu^{\prime}|)!}},
\end{array}
\label{eq.app.2.7}
\end{equation}

\vspace{-4mm}
\begin{equation}
\begin{array}{lcl}
\vspace{1mm}
  C_{l_{i} l_{f} l_{\gamma} n}^{m_{i} \mu} & = &
  (-1)^{l_{f} + n + m_{i} + |m_{i}|}\;
  i^{l_{f}+l_{i} + n} \cdot
  (n, 1, l_{\gamma} \,\big| \,0, \mu, \mu) \cdot
    \sqrt{
      \displaystyle\frac{(2l_{f}+1)\, (2l_{i}+1)\, (2n+1)}{16\pi}
      \displaystyle\frac{(l_{f}-|m_{i}|)!}{(l_{f}+|m_{i}|)!}\;
      \displaystyle\frac{(l_{i}-|m_{i}|)!}{(l_{i}+|m_{i}|)!} },
\end{array}
\label{eq.app.2.8}
\end{equation}

\vspace{-5mm}
\begin{equation}
  f_{l_{1} l_{f} l_{2}}^{m_{i} m_{f} \mu^{\prime}}(\theta) =
    P_{l_{1}}^{|m_{i} - \mu^{\prime}|}(\cos{\theta})\;
    P_{l_{f}}^{|m_{f}|}(\cos{\theta})\;
    P_{l_{2}}^{|m_{i} - m_{f} - \mu^{\prime}|} (\cos{\theta}).
\label{eq.app.2.9}
\end{equation}
We define differential functions on the integrals (\ref{eq.app.2.6}) with angular dependence as
\begin{equation}
\begin{array}{lcl}
  \displaystyle\frac{d\, I_{M}\, (l_{i}, l_{f}, l_{\gamma}, l_{1}, \mu)} {\sin{\theta}\,d\theta} & = &
  \delta_{\mu, m_{i}-m_{f}}\;
  \displaystyle\sum\limits_{\mu^{\prime} = \pm 1}
    C_{l_{i} l_{f} l_{ph} l_{1} l_{ph}}^{m_{i} m_{f} \mu^{\prime}} \cdot
    f_{l_{1} l_{f} l_{\gamma}}^{m_{i} m_{f} \mu^{\prime}}(\theta), \\

  \displaystyle\frac{d\, I_{E}\, (l_{i}, l_{f}, l_{\gamma}, l_{1}, l_{2}, \mu)} {\sin{\theta}\,d\theta} & = &
  \delta_{\mu, m_{i}-m_{f}}\;
  \displaystyle\sum\limits_{\mu^{\prime} = \pm 1}
    C_{l_{i} l_{f} l_{ph} l_{1} l_{2}}^{m_{i} m_{f} \mu^{\prime}} \cdot
    f_{l_{1} l_{f} l_{2}}^{m_{i} m_{f} \mu^{\prime}}(\theta), \\

  \displaystyle\frac{d\, \tilde{I}\, (l_{i}, l_{f}, l_{\gamma}, n, \mu)} {\sin{\theta}\,d\theta} & = &
    \delta_{m_{i} m_{f}}\,
    C_{l_{i} l_{f} l_{\gamma} n}^{m_{i} \mu}
    f_{l_{i} l_{f} n}^{m_{i} m_{i} 0}(\theta).
\end{array}
\label{eq.app.2.10}
\end{equation}

\section{Calculations of the wave functions of relative motion between two nuclei
\label{sec.app.4}}

\subsection{Boundary conditions and normalization of the wave functions
\label{sec.app.4.1}}

In this paper for the wave function of relative motion $\phi (\vb{r})$ in the initial $i$-state and final $f$-state we chose states of the elastic scattering of one nucleus on another nucleus,
for which we have used the normalization condition for the radial wave function of relative motion $R_{s}(k,r)$ as
(see Ref.~\cite{Landau.v3.1989}, p.~138)
\begin{equation}
\begin{array}{lll}
  \int\limits_{0}^{+\infty} R_{s}^{*}(k',r)\, R_{s}(k,r)\: r^{2} dr =
  \int\limits_{0}^{+\infty} \chi_{s}^{*}(k',r)\, \chi_{s}(k,r)\: dr = 2\pi\, \delta(k'-k), &
  R_{s}(k,r) = \displaystyle\frac{\chi_{s}(k,r)}{r}.
\end{array}
\label{eq.app.4.1.1}
\end{equation}
Here,
$r$ is relative distance between nuclei,
$k$, $k'$ are wave numbers.
The radial wave function in the asymptotic region can be written as ($s = i, f$)
\begin{equation}
\begin{array}{ccl}
  \chi_{s}(k,r) = N_{s} \; \bigl(A_{s} G_{s}(k,r) + B_{s} F_{s}(k,r) \bigr),
\end{array}
\label{eq.app.4.1.2}
\end{equation}
where $F_{s}$ and $G_{s}$ are the Coulomb functions,
$A_{s}$ and $B_{s}$ are real constants determined concerning the found solution for $\chi_{s}(k,r)$ at small $r$, $N_{s}$ is unknown normalization factor.
At far distances we have
\begin{equation}
\begin{array}{ccl}
  F_{s}(k,r) = \sin{\theta_{s}}, &
  G_{s}(k,r) = \cos{\theta_{s}}, &
  \theta_{s} = k_{s} r - \displaystyle\frac{l\pi}{2} + \sigma_{l}(\eta_{s}), \\
\end{array}
\label{eq.app.4.1.3}
\end{equation}
where $\sigma_{l}(\eta_{s}) = \arg{\; \Gamma(i\eta_{s}+l+1)}$,
$\eta_{s} = \displaystyle\frac{\mu\, m_{N}\, \nu}{k_{s}\, \hbar}$,
$\nu = Z_{1}\, Z_{2}\, e^{2}$ is Coulomb parameter,
$\mu\, m_{N} = m_{1} m_{2} / (m_{1} + m_{2})$ is reduced mass of two nuclei with mass $m_{1}$ and $m_{2}$,
$e$ is electric charge of proton.
At such a representation of the asymptotic Coulomb functions we find
\begin{equation}
\begin{array}{c}
  \vspace{2mm}
  \Bigl(A_{s} G_{s}(k',r) + B_{s} F_{s}(k',r)\Bigr)^{*} \Bigl(A_{s} G_{s}(k,r) + B_{s} F_{s}(k,r)\Bigr) = \\
  =\displaystyle\frac{A_{s}^{2} + B_{s}^{2}}{2} \cos{(\theta'+\theta)} +
  A_{s} B_{s} \sin{(\theta'+\theta)} +
  \displaystyle\frac{A_{s}^{2} + B_{s}^{2}}{2} \cos{(\theta'-\theta)}.
\end{array}
\label{eq.app.4.1.4}
\end{equation}
On such a basis, the integral~(\ref{eq.app.4.1.1}) is transformed to the following form:
\begin{equation}
\begin{array}{c}
    |N_{s}|^{2} \displaystyle\frac{A_{s}^{2} + B_{s}^{2}}{2} \int\limits_{0}^{+\infty} \cos{(\theta'-\theta)} \;dr =
    2\pi \; \delta(k'-k).
\end{array}
\label{eq.app.4.1.5}
\end{equation}
Taking definition of the $\delta$-function into account, we obtain
\begin{equation}
  N_{s} = \displaystyle\frac{2}{\sqrt{A_{s}^{2} + B_{s}^{2}}}.
\label{eq.app.4.1.6}
\end{equation}

In bremsstrahlung problems another normalization condition is useful for decay of nuclear system, so we add this formalism.
The wave function should correspond to emission of cluster from nucleus during unit of time (see Ref.~\cite{Landau.v3.1989}, p.~140):
\begin{equation}
\begin {array} {ll}
  \displaystyle\oint j_{s}\,(\mathbf{\rm r})\; r^{2}\: d\Omega = 1, &
  \chi_{s}(r \to +\infty) = N_{s} \cdot (G_{s}(r)+iF_{s}(r)),
\end {array}
\label{eq.app.4.2.1}
\end{equation}
where $d\Omega = \sin{\theta} \: d\theta \: d\phi$ is solid angle element, $j_{s}({\bf r})$ is probability flux density and the integration is performed over spherical surface of enough large radius, $R_{\rm max}$.
Defining the flux as
$j_{s}\, (\mathbf{\rm r}) = i/2m\: (\phi_{s}({\bf r})\: \nabla \phi_{s}^{*}({\bf r}) - {\rm c.c.})$
%
and choosing the radial component, $\chi_{s}(r)$, at far distances as the outgoing Coulomb wave (see second formula in Eqs.~(\ref{eq.app.4.2.1})),
where $N_{s}$ is unknown normalization factor.
We integrate Eq.~(\ref{eq.app.4.2.1}) over the angular variables and find $N_{s}$ as
\begin{equation}
\begin{array}{ccc}
  \biggl(
    \displaystyle\frac{k_{s}}{m} \cdot \Bigl|\chi_{s}(R_{\rm max})\Bigr|^{2} =
   \displaystyle\frac{k_{s}}{m} \: |N_{s}|^{2} \Bigl(|G_{s}(R_{\rm max})|^{2} + |F_{s}(R_{\rm max})|^{2} \Bigr) =
    \displaystyle\frac{k_{s}}{m} \cdot |N_{s}|^{2} = 1
  \biggr)
  & \to &
  \biggl(
    N_{s} = \sqrt{\displaystyle\frac{m}{k_{s}}} \;
  \biggr).
\end{array}
\label{eq.app.4.2.2}
\end{equation}

\subsection{Aspects of numerical calculations of the wave functions
\label{sec.app.4.3}}

To calculate integrals (\ref{eq.2.4.6.3}) and (\ref{eq.2.4.6.6}),
we used large but finite range for the variable $r$: $0 \le r \le R_{\rm max}$.
We separate the full radial region on the internal and asymptotic parts at point $R_{\rm at}$:
The internal region with strong effects of nuclear and Coulomb interactions between clusters ($0 \le r \le R_{at}$), and
the asymptotic region with the Coulomb interaction only ($R_{\rm at} \le r \le R_{\rm max}$).
Calculations of the wave functions are performed in each region independently by different methods.

In the internal region we use the following way.
For the state of elastic scattering the wave function, $\chi_{s}$, is real.
We determine each partial solution of wave function and its derivative at a selected starting point $r_{0}$, and
then we calculate those in the region close enough to this point using the \emph{method of beginning of the solution} (based on expansion on Teylor series, see Ref.~\cite{Kunz.1964.book}).
For the solution increasing in the barrier region we choose $r_{0}=0$ as the starting point.
For the solution decreasing in the barrier region we choose $r_{0} = R_{\rm at}$ as the starting point.
Then we calculate both partial solutions and their derivatives independently in the full nuclear region using the \emph{method of continuation of the solution}
shortly presented in App.~B.3 in Ref.~\cite{Maydanyuk.2011.JPG},
which is improvement of the Numerov method with constant step \cite{Kunz.1964.book}.
Then, we find the unknown complex coefficients from the corresponding boundary conditions.

The Coulomb wave functions and their derivatives in the asymptotic region are calculated by using library programs.
Then, they are matched at point $R_{\rm at}$ with the solutions in the internal region, using continuity conditions for the wave functions and their derivatives.
The $R_{\rm max}$ boundary is chosen from requirement to achieve convenient stability and convergence in the calculations of the cross sections
(this can be boundary about 200000~fm).


\end{document}